# Enhanced $^{133}Cs$ Triple-Quantum Excitation in solid-state NMR of Cs-bearing Zeolites

N. Vaisleib[1], M. Arbel-Haddad[2], A. Goldbourt[1§]

[1]School of Chemistry, Tel Aviv University, Ramat Aviv 6997801, Tel Aviv, Israel
[2]Nuclear Research Center Negev, PO Box 9001, Beer Sheva 84901, Israel

[§]amirgo@tauex.tau.ac.il

## ABSTRACT

Geopolymers are aluminosilicate materials that exhibit effective immobilization properties for low-level radioactive nuclear waste, and more specifically for the immobilization of radioactive cesium. The identification of the cesium-binding sites and their distribution between the different phases making up the geopolymeric matrix can be obtained using solid-state NMR measurements of the quadrupolar spin $^{133}Cs$, which is a surrogate for the radioactive cesium species present in nuclear waste streams. For quadrupolar nuclei, acquiring two-dimensional multiple-quantum experiments allows the acquisition of more dispersed spectra when multiple sites overlap. However, $^{133}Cs$ has a spin-7/2 and one of the smallest quadrupole moments, making multiple-quantum excitation highly challenging. In this work we present pulse schemes that enhance the excitation efficiency of $^{133}Cs$ triple quantum coherences by a factor of ~2 with respect to a two-pulse excitation scheme. The improved schemes were developed by using numerical simulation and verified experimentally by applying one and two-dimensional triple-quantum solid-state NMR experiments to a mixture of cesium-exchanged hydrated zeolites A and X, which possess dynamically averaged small quadrupolar coupling constants in the order of 10 kHz.

## 1. INTRODUCTION

Multiple-quantum solid-state NMR holds considerable promise for investigating complex molecular structures due to the increase in spectral dispersion with respect to direct detection and elevated coupling values to surrounding spins. However, its application is often limited by low excitation efficiency of triple-quantum and higher-quantum coherences, resulting in low signal to noise ratios. Exciting multiple-quantum coherences in nuclei with weak quadrupolar coupling constants is all the more challenging because weak quadrupolar interaction results in smaller energy mixing of the Zeeman states.

Few nuclei in the periodic table have a quadrupolar spin with a natural low quadrupolar moment, the main examples being $^{2}H$, $^{6}Li$, and $^{133}Cs$. The latter is a spin-7/2 alkali metal that can be found in its ionic form in industrially important materials such as perovskites for solar energy applications[1], organic-free zeolites for carbon capture[2,3] and for catalysis[4,5]. Studies of Cs adsorption to zeolites and similar aluminosilicate materials have been conducted in the context of radioactive waste treatment, where $^{133}Cs$ is often used as a surrogate for the radioactive cesium species present in nuclear waste streams [6]. These are by-products of nuclear operations such as power generation, research and medicine. Sequestration of the radioactive species in these waste streams is necessary in order to prevent their release to the environment. Cement-based materials, including ordinary Portland cement (OPC), are commonly used for solidification and sequestration of low and intermediate-level radioactive wastes[7–9]. However, cesium is highly soluble in the alkaline environment afforded by the cement matrix, and is therefore not efficiently immobilized by cement based systems[8,10]. Immobilization of radioactive wastes by geopolymers, which are materials obtained through the hydrothermal reaction between an aluminosilicate source and an alkaline reagent, has been suggested as an affordable and efficient alternative for nuclear waste immobilization and sequestration, and more specifically for Cs-bearing nuclear wastes [10,11].

Geopolymers are inorganic materials with a three-dimensional Si-O-Al polymeric network[12]. The chemical nature of geopolymers is similar to that of zeolites[13]. In both cases the aluminosilicate framework is negatively charged due to the presence of tetrahedrally bonded aluminate sites. This negative charge is balanced by the presence of cations within the structure[13,14]. However, while zeolites are well ordered crystalline materials, geopolymers may be either fully amorphous or contain both amorphous and crystalline domains[13,15]. Geopolymers have been shown to have chemical durability similar to that of zeolites[16,17], and they possess many properties which are desirable for waste immobilizing applications, such as heat and radiation resistance, and high mechanical strength, in addition to their strong affinity for the waste cations [15,17].

The characterization of geopolymers is often achieved by combining X-ray diffraction (XRD), FTIR spectroscopy, porosity and surface area measurements, as well as solid-state NMR[18]. Most solid state NMR studies of geopolymers have focused on characterization of the aluminosilicate structure[19–22]. Fewer studies report on the binding sites and binding modes of cations within this framework[23]. For cesium cations, which are the focus of interest in the case of nuclear waste immobilization, only few such examples exist for geopolymers[24]. We have recently shown that by combining data from XRD and $^{133}$Cs solid-state NMR it is possible to distinguish between Cs ions in the different phases of geopolymeric matrices and to follow the leaching process in a phase-by-phase manner[25]. Yet, our data indicated that full spectral dispersion is challenging given the natural heterogeneity of such systems. While the most sensitive NMR modality allowing site distinction is the chemical shift, the quadrupolar nature of cesium may allow further improvement in site resolution via multiple-quantum techniques. Here we show, through numerical simulations correlated with experimental data, that despite the apparently low quadrupolar coupling constant ($C_q$) of $^{133}$Cs in hydrated zeolites, improvement in the ability to excite triple-quantum coherences (TQC) in cesium can be facilitated by multiple-pulse techniques. The model system chosen in order to demonstrate this is a mixture of two Cs-exchanged zeolites, zeolite A and zeolite X, which were identified within the Cs-binding geopolymers studied in our previous studies[25,26].

## 2. MATERIALS AND METHODS

### 2.1 Sample Preparation

Zeolite 4A (ZA, $Na_2O:Al_2O_3:2SiO_2:4H_2O$) and zeolite 13X (ZX, $Na_2O:Al_2O_3:2.8SiO_2:4H_2O$) were purchased from Sigma-Aldrich in the Na-form. Partial exchange of $Na^+$ ions by $Cs^+$ ions was obtained by contacting 1g samples of the respective zeolites in 40mL of $CsCl_{(aq)}$ solution (0.1M). The ion-exchange process reached equilibrium in less than 24 hours. The concentration of both Na and Cs ions in the solution at equilibrium was determined by ion chromatography (IC), as described previously[25]. This data was used to calculate the fraction of $Na^+$ ions exchanged by $Cs^+$ ions, which was found to be 0.6 in ZA and 0.75 in ZX. A mixture of the two Cs-loaded zeolites was used for NMR measurements.

### 2.2 Simulations

All computer simulations were conducted using the SIMPSON[27] software and data were further analyzed by home-written python scripts, which are available on GitHub (https://github.com/noyvais/Master/tree/main).

### 2.3 Solid State NMR Experiments

Solid-state NMR experiments were carried out on a Bruker Avance III spectrometer operating at $^{133}$Cs resonance frequency of 78.7 MHz (magnetic field of 14.0 T) and equipped with a triple-resonance 4 mm magic-angle spinning probe set to double-resonance mode.

All experiments were carried out in room temperature without employing $^1$H decoupling (proven to be unnecessary under MAS). The chemical shifts were externally referenced to $CsCl_{(s)}$ at 223.1 ppm. The radiofrequency (RF) power $\nu_1$ that was used for all $^{133}$Cs NMR experiments was determined from nutation experiment on $CsCl_{(s)}$ to be $\nu_1$=37.5 kHz (350 W, the limit of our probe). Relaxation delays were set to 0.5 seconds, calculated from saturation recovery to be ~3 times the maximal $T_1$ value in the sample. Other experimental parameters are provided in the captions.

## 3 RESULTS AND DISCUSSION

### 3.1 The Quadrupolar Coupling Constant in Hydrated Zeolites A and X

Prior studies on CsA (a sample with 30% exchange) suggested that at room temperature the $^{133}$Cs quadrupolar coupling constant $C_q$ is sufficiently small, so that second order effects are undetected[28]. Further evidence arises from the report that $^{133}$Cs nutation spectra are similar to those of Cs in solution[29]. Slow MAS (3.5 kHz, 4.7T) studies of hydrated Cs-exchange zeolite NaX[30] have also revealed narrow lines suggesting a small Cq. In order to obtain an estimated upper limit for the quadrupolar coupling constant in our specific model zeolites, we carried out nutation experiments, lineshape analysis under static and MAS conditions and simulations at the relevant experimental conditions. In CsA, it was suggested that line broadening at low temperature is a result of the dipolar interaction with the surrounding protons[29]. Here we examine the lineshapes at room temperature also considering broadening from quadrupolar satellites.

### 3.1.1 Nutation experiments:

For quadrupolar nuclei, the nutation frequency $\nu_{nut}$ depends strongly on the nuclear quadrupolar frequency $\nu_q$[31,32], which for a spin I=7/2 is related to the quadrupolar coupling constant $C_q$ by $\nu_q = \frac{3C_q}{2I(2I-1)} = \frac{C_q}{14}$. The nutation frequency is given by

$$(1)\ \nu_{nut} = \begin{cases} \left(I + \frac{1}{2}\right)\nu_1 & \text{for} \quad \nu_q \gg \nu_1 \\ \nu_1 & \text{for} \quad \nu_q \ll \nu_1 \end{cases}$$

In the intermediate regime ($\nu_q \cong \nu_1$), the nutation frequency is in between the values of Eq. (1). An example of simulations showing such behavior for spins 3/2 and 5/2 can be found[33]. For $^{133}$Cs, a spin-7/2, the simulations in figure 1 reproduce and somewhat extend those previously shown[34,35]. They show an increase in $\nu_{nut}$ alongside the decrease in the central-transition (CT) intensity as $\nu_q$ increases with respect to $\nu_1$. When $\nu_q \gg \nu_1$, the signal keeps decaying due to transfer of magnetization to additional spin states. Using the nutation profile, it is possible to determine bounds to the value of $\nu_q$ even for sites that are characterized by broad lines without clear second order patterns.

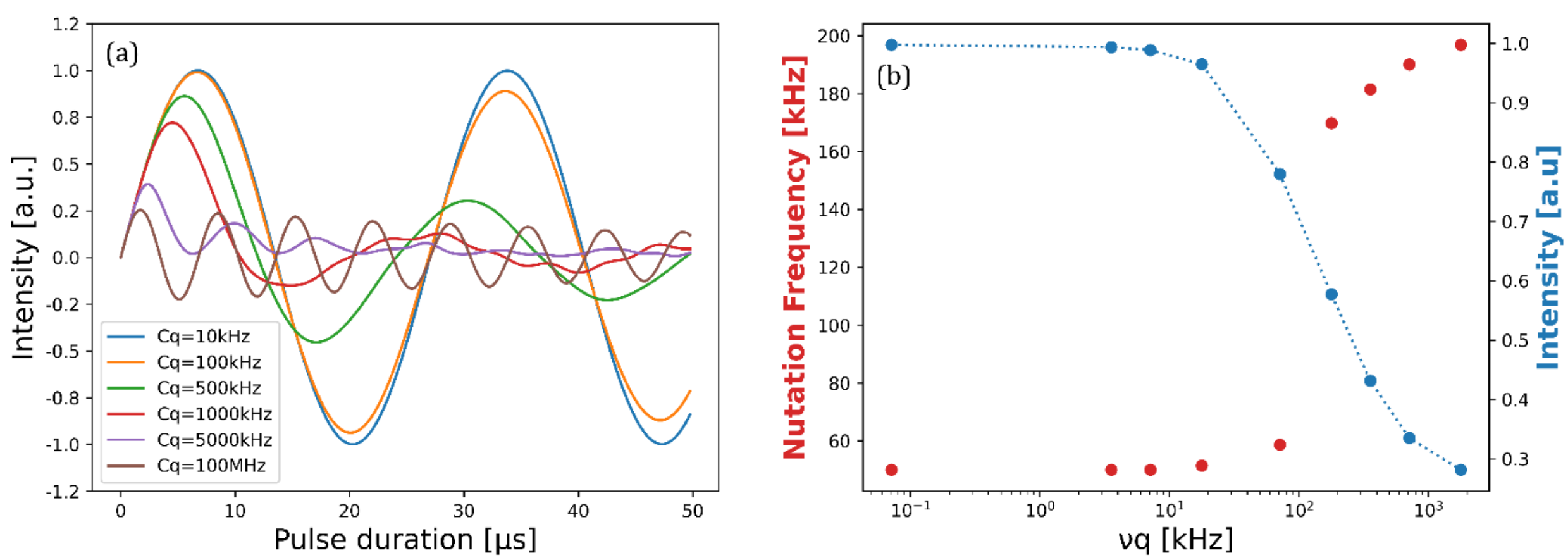

***Figure 1:*** *(a) Simulated single-quantum central-transition (CT) $^{133}$Cs NMR signal as a function of the pulse duration for different $C_q$ values. In these simulations, we used $\nu_1$=50 kHz, $\nu_R$=5 kHz and CT detection. The intensity is normalized to a CT signal obtained after a 90° pulse. (b) Nutation frequency and signal intensity as a function of $\nu_q$. The nutation frequencies were determined by the first zero crossing of the different plots presented in (a).*

Figure 2 shows the experimental CT $^{133}$Cs nutation spectra of the mixed zeolites' sample. It shows that Cs in both ZX (left, larger peak in the doublets) and ZA (right peak) have a similar nutation frequency, $\nu_{nut}$ = 42.3 kHz $\cong \nu_1$, implying that the $C_q$ values of $^{133}$Cs in both zeolites are small relative to the RF field intensity ($\nu_q \ll \nu_1$) and are of the same order of magnitude.

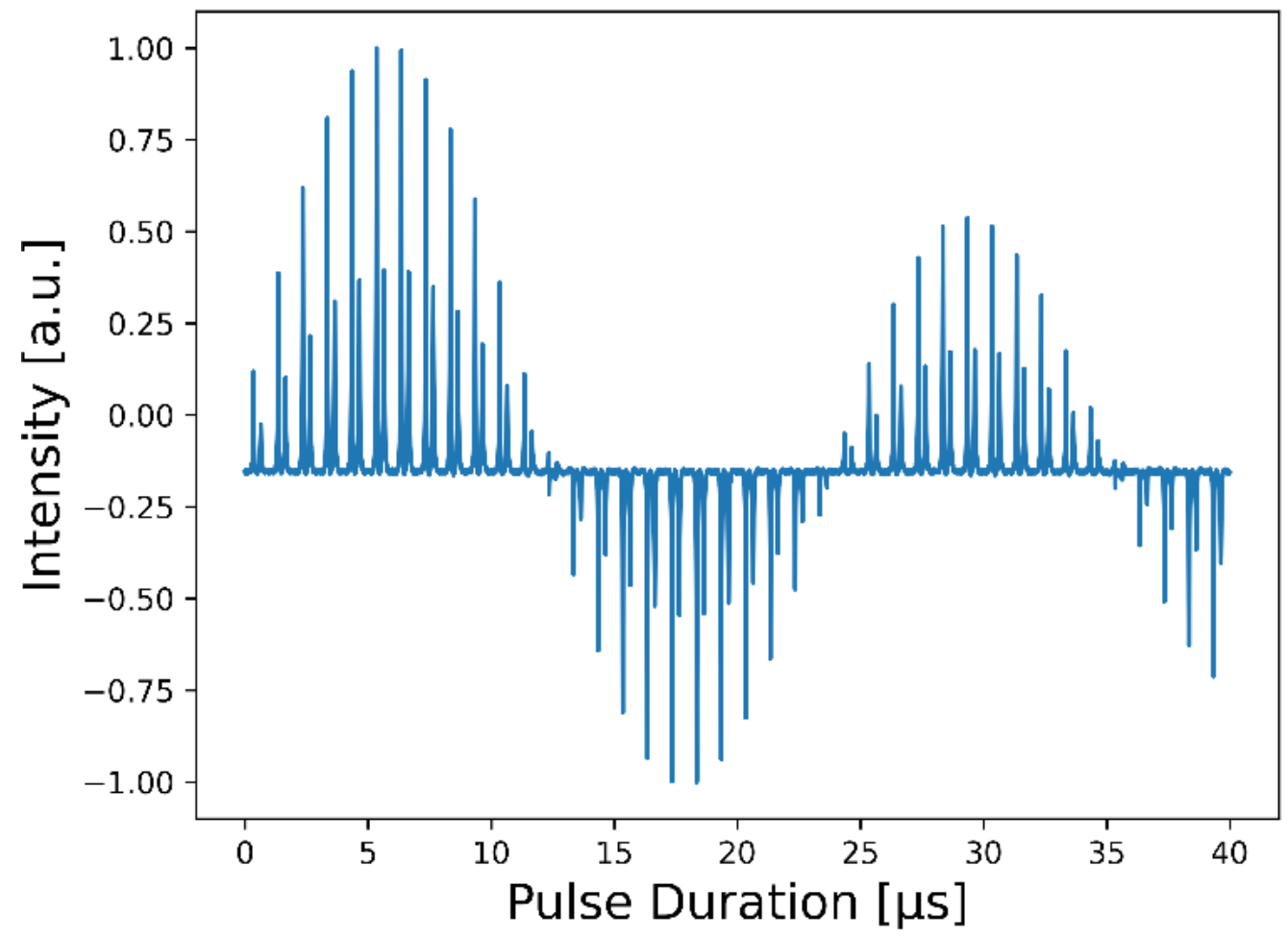

***Figure 2:*** *Experimental* [133]*Cs SQ nutation spectra on a sample containing zeolite X (left, larger peak in the doublet) and zeolite A (right peak). We used* $\nu_1$*=37 kHz,* $\nu_R$*=5 kHz and 8 scans. The average of the three zero crossings gives* $\nu_{nut}$*=42.3 kHz. The two sites were assigned by measuring samples of zeolites A and X separately.*

### 3.1.2 Lineshapes analysis:

The static lineshapes and the spinning sideband intensities under MAS can also be used to estimate the value of $C_q$ in $^{133}$Cs[36–38]. Using a first order approximation, the transition frequency $|m\rangle \rightarrow |m+1\rangle$ due to the quadrupolar interaction is:

$$(2)\ \nu_{q\,shift}^{LAB} = \nu_q \cdot f(\Omega, \omega_r)(m + ½)$$

where $f$ for the static and MAS cases can be found elsewhere[39,40]. While the CT is not shifted in first order, every satellite transition (ST) is shifted by an additional factor of $\pm\nu_q f(\Omega, \omega_r)$, which results in powder broadening in the static case, but under MAS the powder lineshapes of the satellites collapse into multiple spinning sidebands. In systems where the quadrupolar coupling constant is large enough one must consider the second-order correction, which is proportional to $\nu_q^2/\nu_0$. Due to the low electric quadrupole moment of cesium[41] this term can be neglected here.

Without the effects of the second-order quadrupolar broadening, static spectra of quadrupolar nuclei present typical quadrupolar lineshapes, in which only the STs are broadened and the CT is unaffected[40,42]. When the quadrupolar interaction is very small, the overlap of the ST with the CT is manifested in broadening of the central line. Such broadening is depicted in figure 3, where we simulated static and MAS $^{133}$Cs NMR spectra with different $C_q$ values that are commonly observed in cesium compounds[43]. We acquired both static and MAS experimental $^{133}$Cs NMR spectra and can use the resulting lineshapes to estimate the $C_q$ values of each of the two sites ZA and ZX. It should be noted that the chemical shift anisotropy and the dipolar interaction with neighboring protons may also contribute to the broadening of the central transition lineshape.

As shown in figure 3 and in figure S1.1 of the supplementary material, for $C_q$ values under 50 kHz the STs of the static spectra strongly overlap with the CT. When $C_q$=5 kHz, this results in apparent broadening of the CT, similarly for Cq=10 kHz, while when $C_q$=20-50 kHz, ST are somewhat more spread. Here we used an exponential line broadening factor of 100 Hz that still allows distinguishing the sharp discontinuities. However, experimentally even a small heterogeneity in the Cs environment, or enhanced relaxation, can eliminate the sharp features and cause apparent spectral broadening in the base of the peak. In the experimental spectra presented in figure 3(d), the static spectrum is slightly broadened with respect to the MAS spectrum, resembling the simulation in figures 3(a) and S1.1. It suggests that Cq≃5-10 kHz and that any anisotropy of the quadrupolar interaction for both zeolites is very low, in agreement with our nutation results and prior result on Cs-exchanged zeolite A[29]. Partial overlap of the two peaks and additional factors such as heterogeneity, dynamics, chemical shift anisotropy (CSA), as well as dipolar coupling with protons of surrounding water molecules (estimated to be one or two molecules) may also affect the lineshapes[28]. Dipolar broadening has been observed at low temperatures, and at room temperature conditions, as we used here, it is reasonable to assume that the water protons are highly dynamic. Overall, our lineshape analysis estimates of the quadrupolar coupling constants are upper bound values and if any dipolar broadening still exists, the $C_q$ estimates will be further reduced.

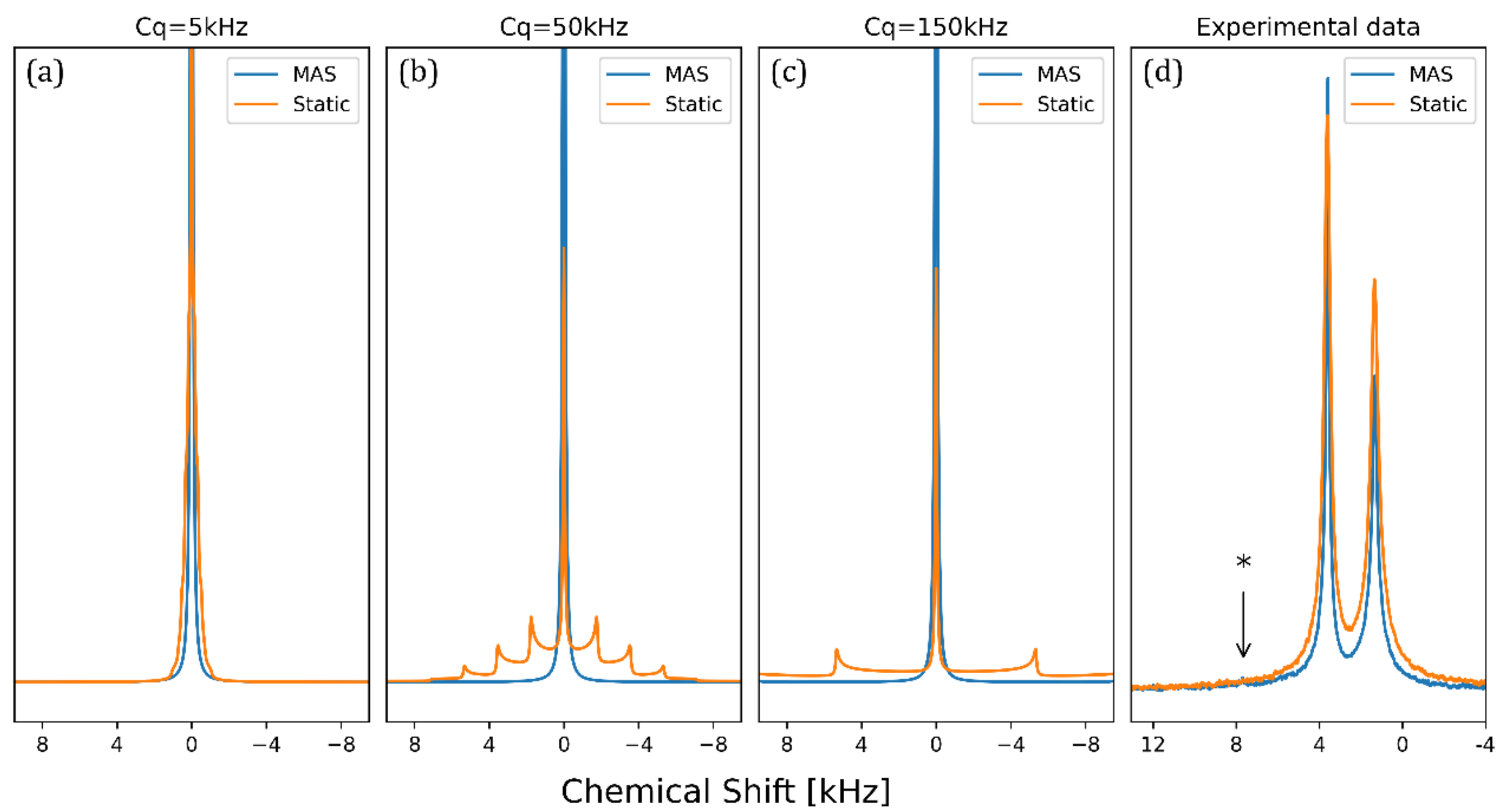

***Figure 3:*** *(a-c) Simulations of* $^{133}$*Cs MAS (blue,* $\nu_R$*=15 kHz) and static (orange) lineshapes for different* $C_q$ *values, using* $\eta_q$*=0. Only the first-order quadrupolar interaction was considered. In order to include ST sidebands, the detection operator was set to* $I^+$*. Line broadening of 100 Hz was applied in all cases. In (c), two of the three STs are out of the spectral region demonstrated in this figure. (d)* $^{133}$*Cs solid-state NMR experimental lineshapes. Line broadening of 30 Hz was applied to both MAS and static spectra, and in the MAS spectrum the spinning rate was* $\nu_R$*=5 kHz. The scale of the static spectrum was magnified by a factor of 1.4 for clarity. A single spinning sideband is indicated by the arrow. The number of scans was 8.*

Another means to estimate $C_q$ is by using the relative intensity R of the first spinning side-band S(ssb) with respect to that of the main peak intensity S(CT). For a given spinning speed, R depends on the anisotropy and the asymmetry parameters of the dominant interaction[44], here assumed to be the quadrupolar interaction. As stated above, it is reasonable to assume that at room temperature the dipolar interaction and CSA are also dynamically averaged and are very small. The broadening of the static spectra is significantly smaller than the spinning speed and therefore if its origin is from CSA or dipolar couplings, spinning sidebands are unlikely to be observed. The quadrupolar satellites on the other hand can span higher frequencies as they are present up to 3 times the quadrupolar frequency for a given orientation a spin-7/2 (Eq. 2). Figure 4 shows simulated MAS spectra with different $C_q$ values, and the corresponding value of R. These are then compared to the experimental result shown in figure 3(d), for which R≈0.017. According to our simulations, a $C_q$ value of ~20 kHz is the closest to the experiment. While this low value is again in the same order of magnitude as we observed by nutation experiments and lineshape analysis, it is still an estimate, due to the possible existence of additional interactions, and since the unknown asymmetry parameter can also affect this value. Additionally, spectral deconvolution of the experimental data, which may manifest some site heterogeneity or exchange, with such small spinning sidebands, produces insufficiently accurate lineshapes. Thus, we can only narrow down the range of $C_q$ values and not determine the $C_q$ value precisely. Yet, the estimates we obtain for $C_q$ are sufficient to guide our efforts to generate TQC with improved efficiency.

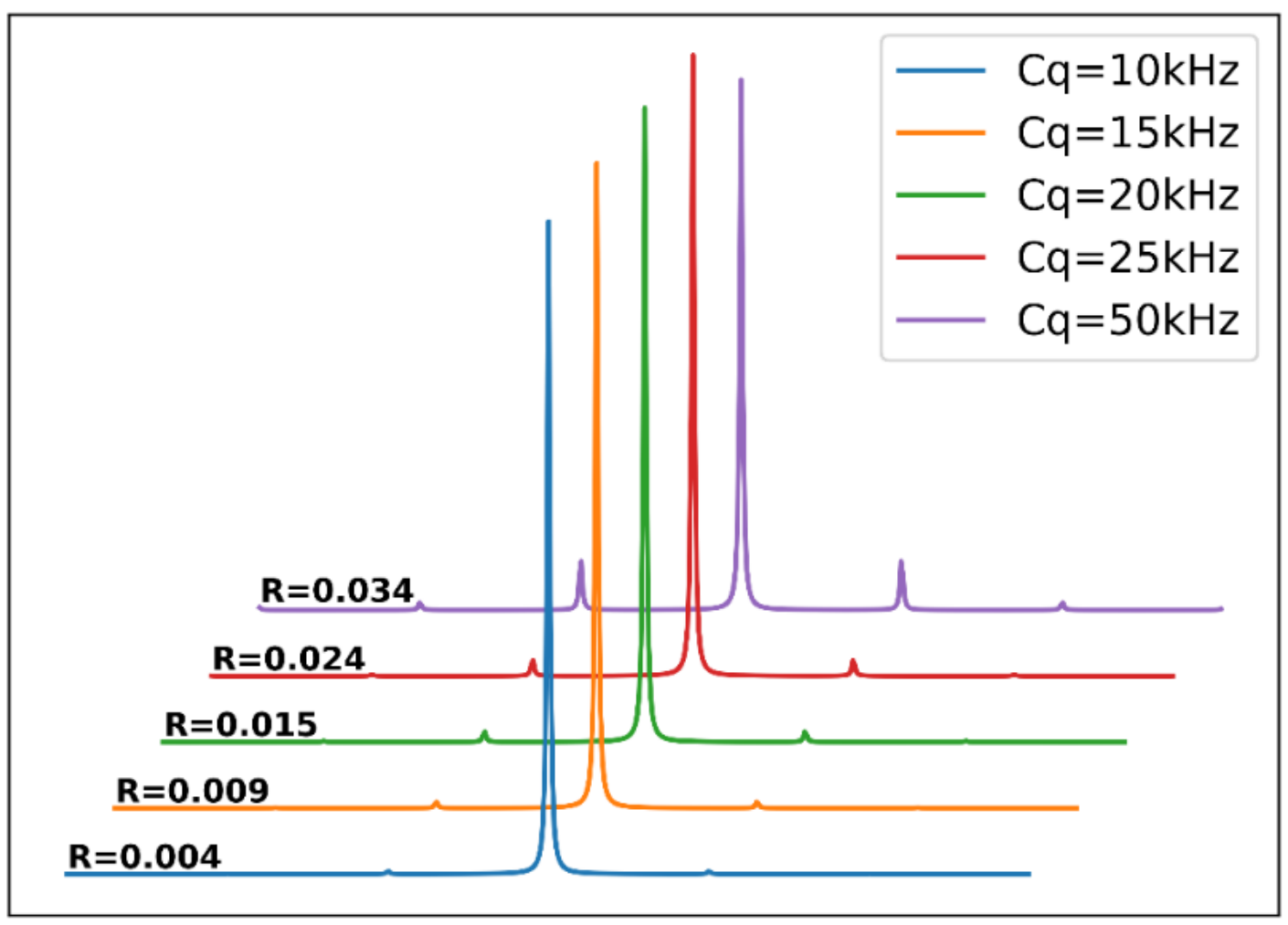

**Figure 4:** *Simulations of $^{133}$Cs MAS spectra for different $C_q$ values. R is the ssb/CT-peak ratio defined in the text. In all simulations, η=0.12 and the spinning speed is 5 kHz.*

## 3.2 Triple-Quantum Spectroscopy – Simulations

When studying materials with multiple Cs binding sites using $^{133}$Cs NMR, overlapping and broad peaks may be observed, e.g. our recent study of cesium leaching in geopolymers[25], making it difficult to analyze the spectra. It is well known that the use of triple-quantum (TQ) spectroscopy enables separation of closely spaced signals and should therefore assist in the study of such systems. Another advantage of TQ spectroscopy is the enhancement of the dipolar interaction thereby improving sensitivity to weakly dipolar-coupled spins. This effect becomes useful in correlation experiments[45] and enables spectral editing[46]. It will therefore be highly useful for examining the coordination and binding of cesium cations in zeolites. Here we focus on the first step of achieving enhanced sensitivity when $C_q$ is very small.

### 3.2.1 Single-Pulse Excitation of TQC

Figure 5 shows simulated $^{133}Cs$ TQ nutation curves and allows us to find the best conditions for a single-pulse excitation of TQCs for different $C_q$ values. As we can see, at experimentally realistic RF conditions (for 4mm probes, up to ~50 kHz), low $C_q$ values ($C_q \leq 50$ kHz) respond poorly to a single-pulse excitation, while $C_q$ values in the range of 250 kHz – 1 MHz responded best. However, for $C_q$ values higher than 250 kHz there is a decline in the maximal TQ signal that can be achieved using a one-pulse excitation. Past studies have shown that when $C_q$ is large, multiple-pulse techniques[47–52] and other pulse shaping approaches[53,54] are useful for transferring polarization between single quantum and multiple-quantum coherences.

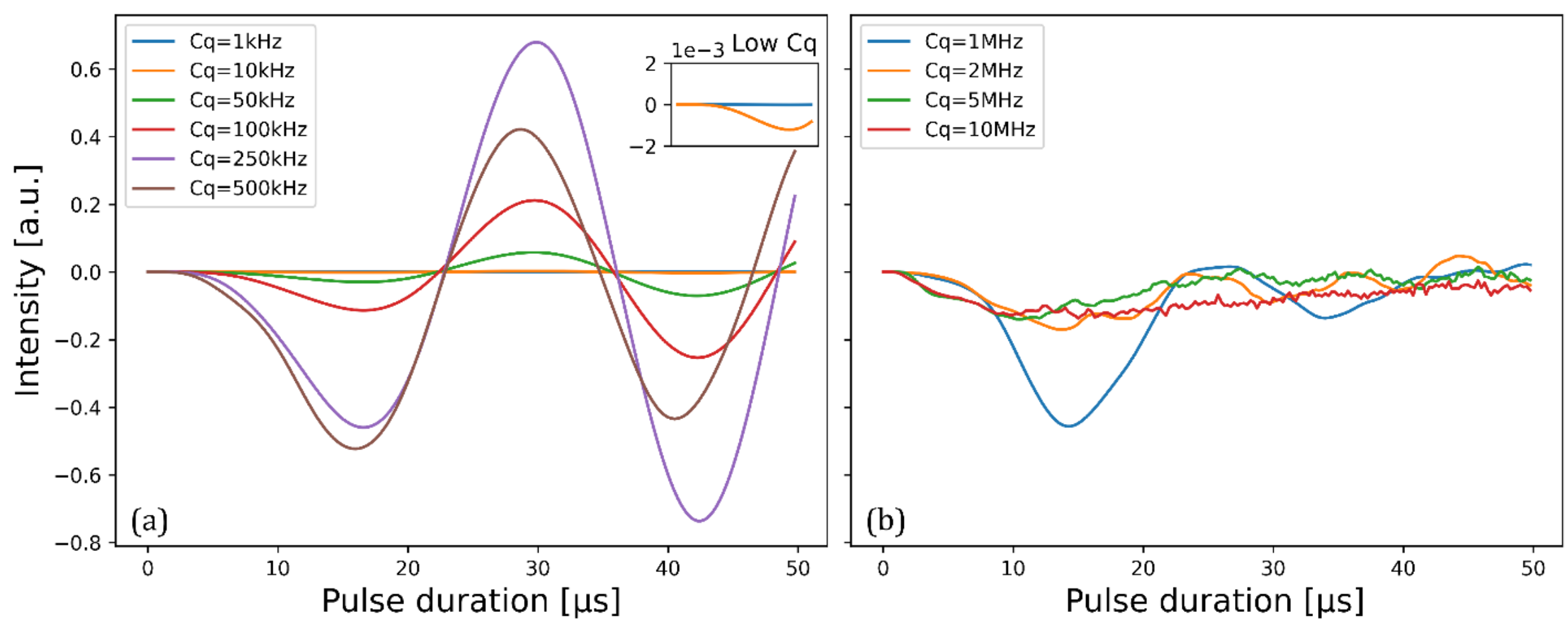

***Figure 5:*** *Simulated intensity of $^{133}Cs$ TQ signal as a function of the pulse duration for different ranges of $C_q$ values. (a) Shows the lower $C_q$ values and (b) shows the higher $C_q$ values. We used $\nu_1$ =50 kHz, $\nu_R$=10 kHz and detection of the triple quantum coherence (the sum of all TQ elements in the density matrix). The intensity is normalized to an ideal CT signal.*

### 3.2.2 Multi-Pulse Excitation of Triple-Quantum Coherences

To further improve excitation of $^{133}Cs$ TQC when $C_q$ is small, we investigated different excitation schemes. Past studies have shown that for a spin-3/2 (e.g., $^7Li$), two short pulses with a delay of one-half rotor period are efficient in exciting TQC[55]. This approach is similar to excitation of multiple-quantum coherences in solution[56]. We therefore focused our attention on pulse blocks where a single pulse is followed by a delay and an additional block containing one to three additional pulses. A reverse excitation scheme (the single pulse is last) has also been examined. Experimentally, only SQC can be detected and therefore both ZQ→TQC and TQC→ZQ blocks (ZQ: zero-quantum coherence, i.e. populations) must be applied, followed by a 90° pulse, as shown in the pulse sequence in figure 6. In simulations,

TQC can be detected directly and therefore examining a ZQ→TQ block is sufficient. In figure 6, each of the pulse block schemes (b) through (g) is used to replace both excitation and conversion blocks in experiment (a), given that both involve a similar coherence transfer pathway.

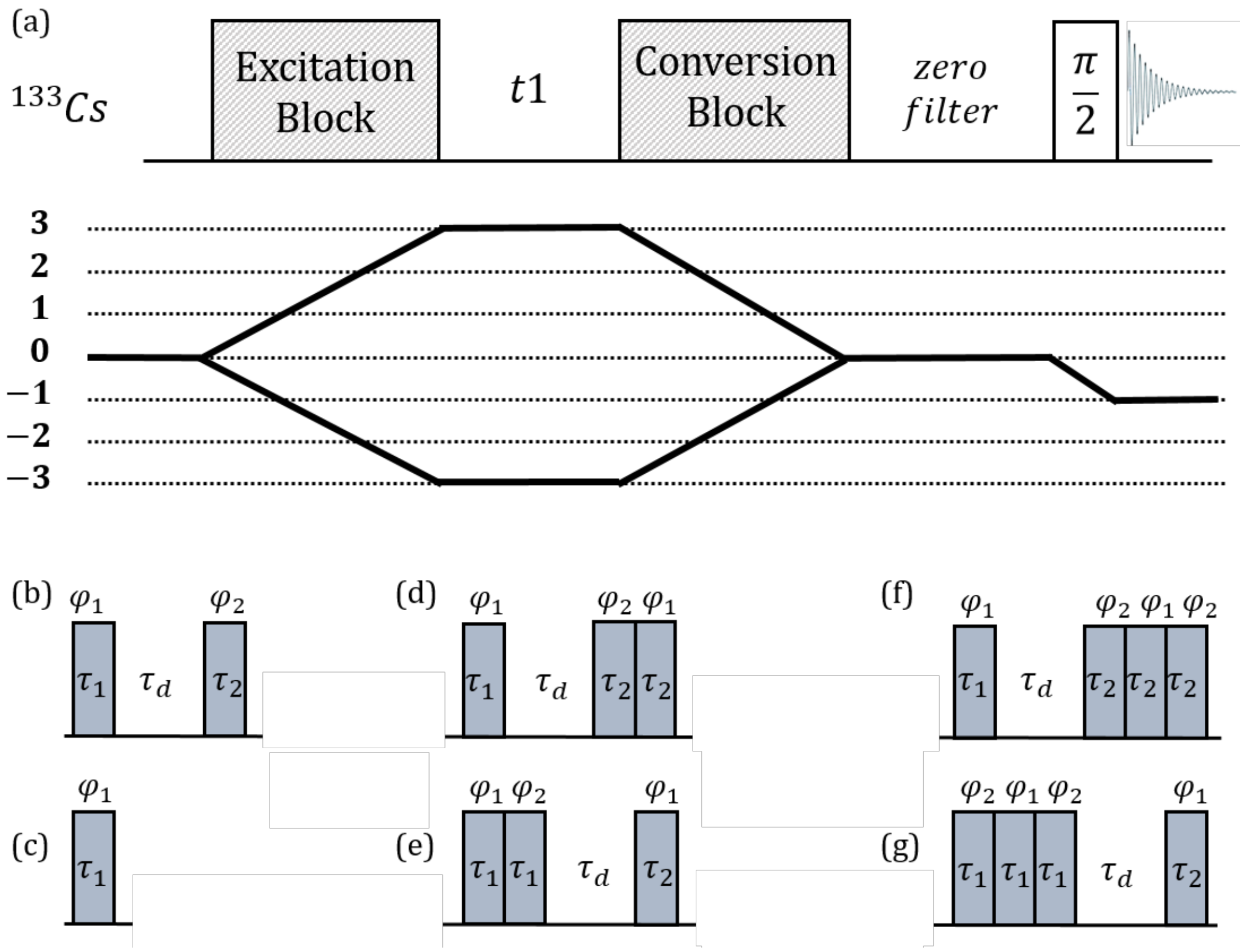

***Figure 6:*** *(a) The scheme used for 1D and 2D TQ-SQ* $^{133}$*Cs correlation experiments, along with the coherence pathway. In the 1D version, t1 evolution time was set to 0.3μs and the TQ filtered signal was detected. Blocks (b) through (g) were used for both excitation and conversion. Block (b) is a two-pulse scheme, (c) is a single-pulse scheme, (d) and (e) are the two options for the three-pulse schemes, and (f) and (g) are four-pulse schemes. Pulse delays are given by* $\tau_d$*, pulse durations by* $\tau_1$ *and* $\tau_2$ *and pulse phases by* $\varphi_1$ *and* $\varphi_2$ *for the excitation block, and* $\varphi_1'$ *and* $\varphi_2'$ *for the conversion block. The full phase cycle used for the pulse sequence was as follows (subscripts on angles are repetitions):* $\varphi_1$ *= 30, 90, 150, 210, 270, 330;* $\varphi_2$ *= 120, 180, 240, 300, 0, 60;* $\varphi_1'=30_6, 90_6, 150_6, 210_6, 270_6, 330_6$*;* $\varphi_2' = 120_6, 180_6, 240_6, 300_6, 0_6, 60_6$*;* $\phi_{\pi/2} = 0$*;* $\phi_{rec} = (0\ 180)_3\ (180\ 0)_3$*. In the optimizations,* $\varphi_1$ *and* $\varphi_2$ *were allowed to interchange.*

Results from the Simpson software simulations of the different excitation schemes are shown in figure 7. In those simulations we applied TQ detection directly. The TQ excitation efficiency depends on the phase of the pulses, the delay between the first and second cluster of pulses and the duration of each pulse. Simulations were performed for different $C_q$ values in the range of 10-450 kHz, four different $\tau_d$ values between ¼ rotor period and a full rotor period, different pulse durations between 1μs and one rotor period (100μs) and different pulse phases. Displayed in figure 7 are the results obtained for pulse durations when $\tau_d$ and the phases of the pulses are optimized. The optimizations performed on $\tau_d$ mostly agree with previous studies on $^{7}$Li[57].

The simulations showed that, in general, the lower the $C_q$ value is, more pulses in the scheme are required to yield better TQC excitation. Even when $C_q$ becomes as low as of 10 kHz, more pulses yielded better excitation, though much weaker than for higher $C_q$ values, and therefore these results are only shown in figure 8. Because $^{133}$Cs compounds tend to have low $C_q$ values[58], we can hypothesize that the best experimental results will be obtained using pulse schemes with a larger number of pulses. Yet here we limit ourselves to a total of four pulses.

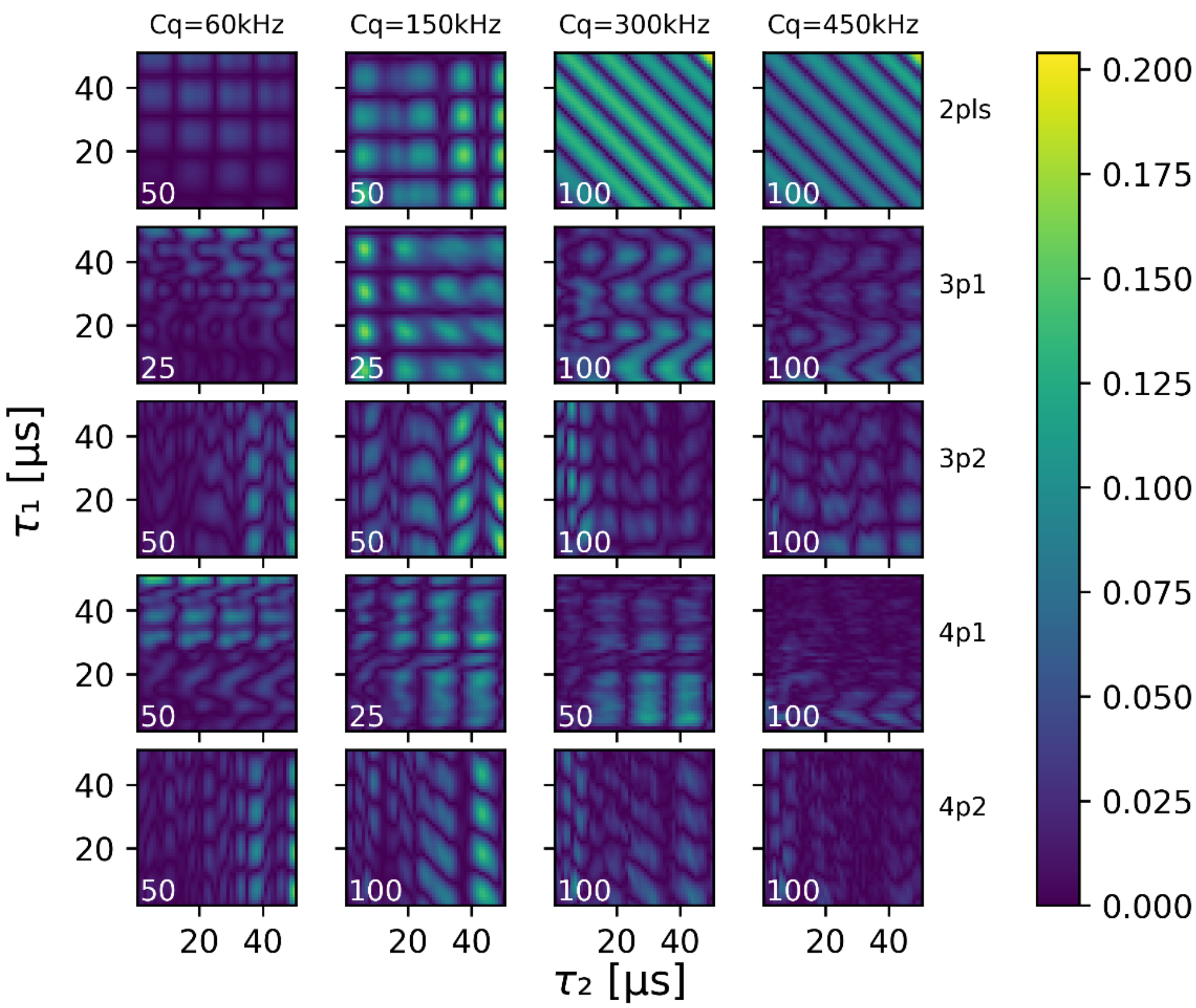

***Figure 7:*** *Heat maps representing the simulated $^{133}$Cs TQ signal using different excitation schemes. In these simulations we applied experimentally realistic values of $\nu_R$=10 kHz and $\nu_1$=40 kHz, and TQ detection was used[1]. The number on the bottom left of each of the plots is the delay between the pulses, $\tau_d$ (in μs). Note that 100μs equals one rotor period. The signal obtained was normalized to an ideal 90-pulse SQ excitation in $^{133}$Cs (detection of $I^+$). The values $\tau_1$ and $\tau_2$ are the pulse lengths as defined in figure 6. In this figure, 2pls corresponds to scheme 6(b), 3p1 to 6(e), 3p2 to 6(d), 4p1 to 6(g) and 4p2 to 6(f).*

[1] In the Simpson software, command "matrix set detect coherence {-3 3}"

Figure 8 summarizes the best TQ signal that could be obtained for each $C_q$, including the value of 10 kHz. When $C_q$>150 kHz ($\nu_q$>21 kHz), the two-pulse scheme, similar to that used for $^7Li$[55], is the most efficient. On the other hand, when $C_q$=60 kHz ($\nu_q$=8.6 kHz), a two-fold enhancement is obtained by the addition of another pulse, and a four-pulse scheme (pulse; delay; three-pulse block) further improves TQ excitation with a two-fold increase in signal when $C_q$=10 kHz ($\nu_q$≈1.5 kHz).

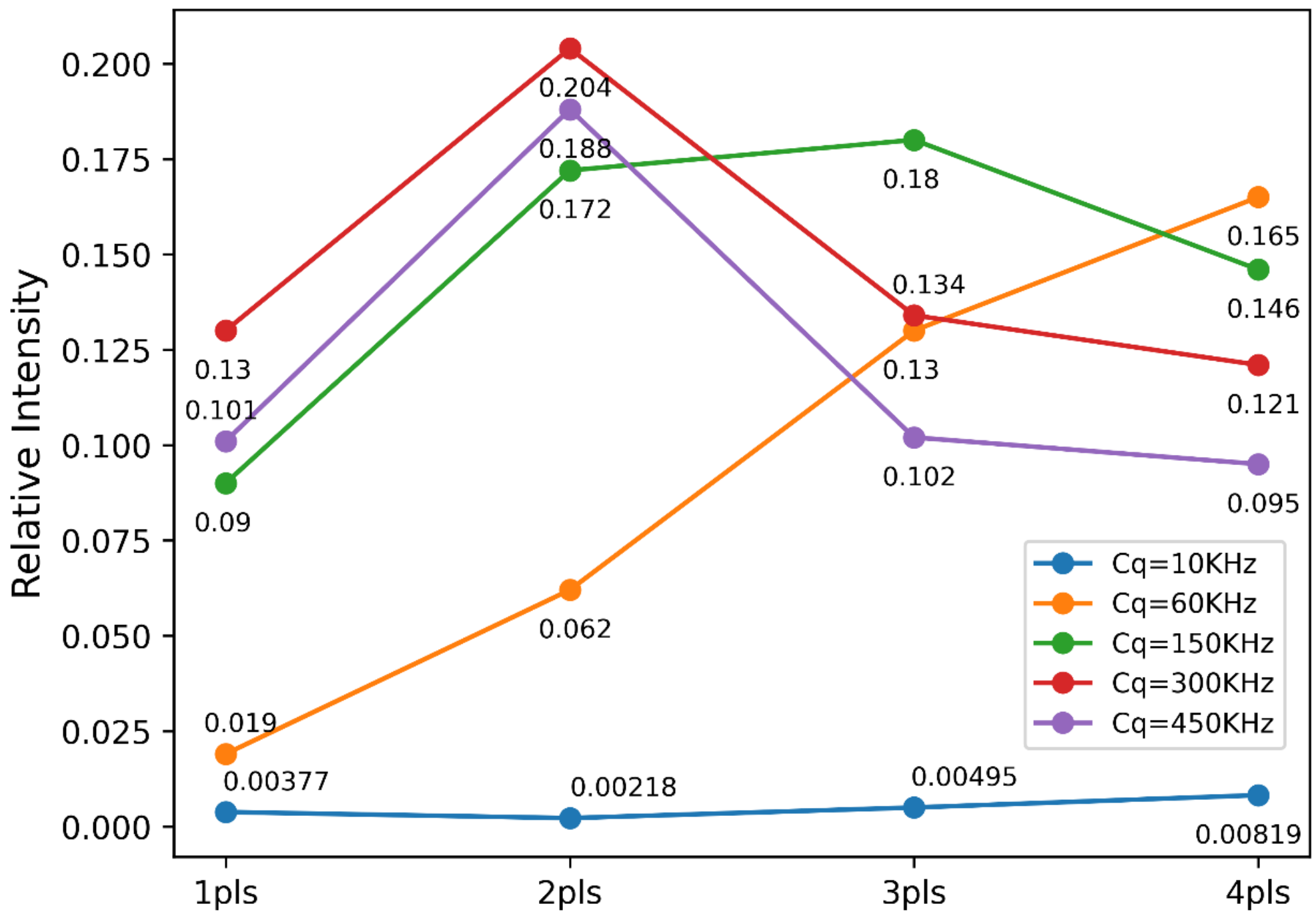

***Figure 8:*** *Maximal TQ signal intensity following a pulse-delay-pulse_block or a pulse_block-delay-pulse excitation scheme under the conditions shown in figure 7. In this figure, 1pls corresponds to the scheme presented in 6(c), 2pls corresponds to 6(b), 3pls corresponds to 6(d, e) and 4pls corresponds to 6(f, g).*

It is clear that the $C_q$ value has a strong effect on the type of the best TQ excitation scheme. To study the effects of the RF power on TQ efficiency, simulations of the schemes presented in figure 6 using different RF power levels were conducted for $C_q$ values of 60-450 kHz and are shown in figure 9. When the pulse power is low ($\nu_1$=10 kHz), the multiple-pulse approach proves efficient. When $\nu_1$=20–70 kHz, the trend shown for 40 kHz is repeated and for high RF power levels $\nu_1 \geq$ 100 kHz, this trend is maintained to some extent and is consistent with the conclusion that for relatively low $C_q$ values, a multiple-pulse scheme is the most efficient approach for TQ excitation. Also noticeable is the fact that if we compare the TQ excitation efficiency of RF values higher than $\nu_1$=70 kHz, we get similar trends and efficiencies suggesting that increasing the RF power beyond 70 kHz does not improve TQ excitation using the pulse schemes of figure 6.

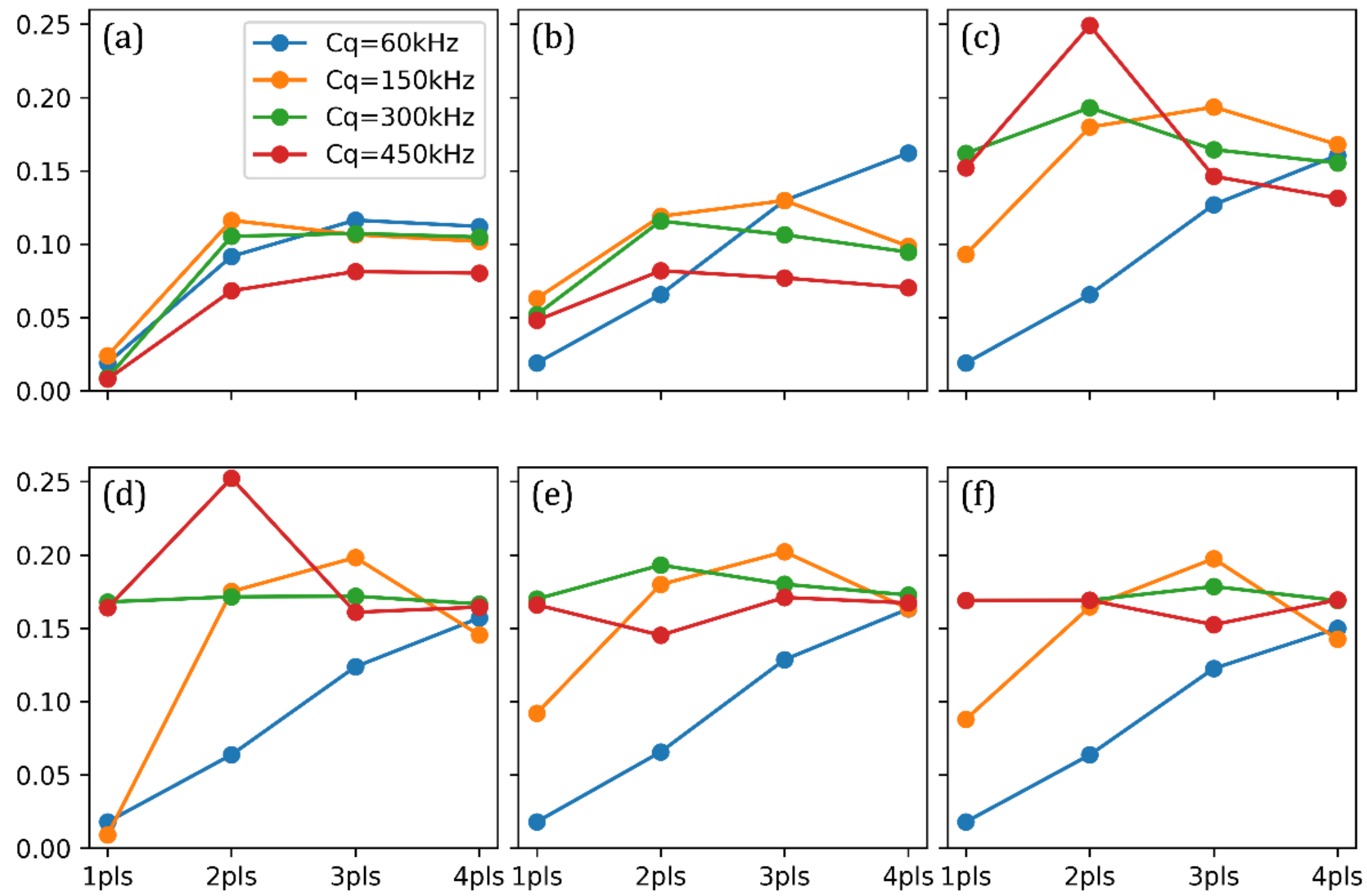

***Figure 9:*** *TQ excitation efficiency as a function of the RF power level $\nu_1$. (a) $\nu_1$=10 kHz, (b) $\nu_1$=20 kHz, (c) $\nu_1$=70 kHz, (d) $\nu_1$=100 kHz, (e) $\nu_1$=150 kHz, (f) $\nu_1$=200 kHz. Similarly to figure 8, 1pls corresponds to the scheme presented in 6(c), 2pls corresponds to 6(b), 3pls corresponds to 6(d, e) and 4pls corresponds to 6(f, g).*

One notable result in our simulations is that long pulses (up to one-half of the rotor period) produce better excitation. However, in the experiments the RF inhomogeneity and relaxation effects can reduce the efficiency of long pulse schemes. We therefore focus in figure 10 on pulse excitation schemes having short, realistically applicable pulse lengths. We also limit ourselves to $^{133}Cs$ sites with $C_q$ values up to 250 kHz. The results indicate once more that for small $C_q$ values ($C_q$ = 10 kHz and 50 kHz), TQ efficiency increases with the number of pulses, while for $C_q$ = 250 kHz TQ efficiency decreases, and a two-pulse scheme is optimal. This means that the trend discussed above is maintained even for shorter pulse lengths.

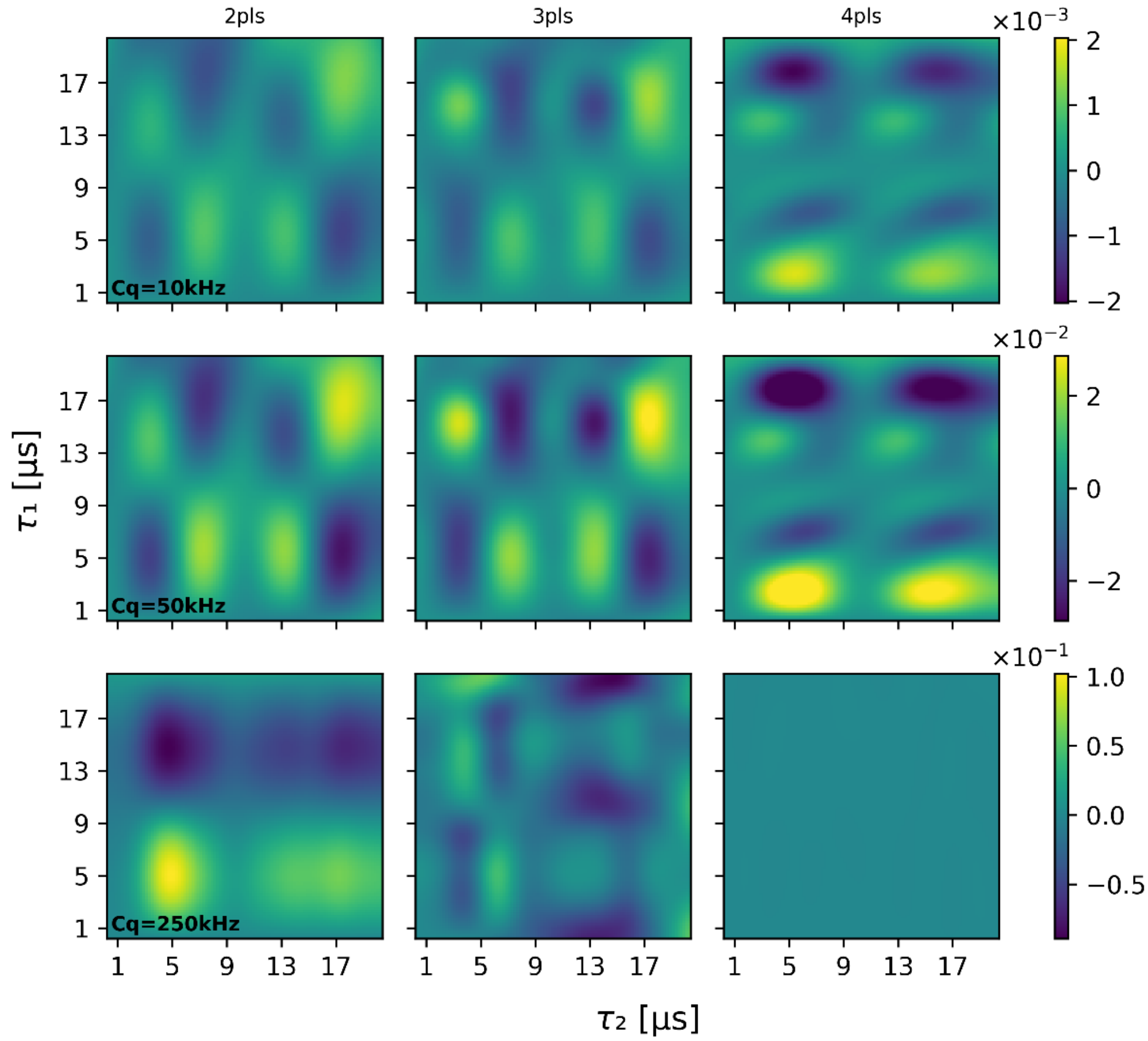

**Figure 10:** *Heat maps representing [133]Cs TQ signal using different excitation schemes at the limit of shorter pulses up to one-quarter of a rotor period (25μs). In these simulations $\nu_R$=10 kHz, $\nu_1$=40 kHz, we detect TQC, and the signal is normalized with respect to <$I^+$> detection resulting from an ideal 90-degree pulse. After optimizations, all delays were set to $\tau_d$=50μs (one-half rotor period). The carrier frequency was chosen to be between the signals of the two zeolites (|Δω|=2 kHz) and the resulting offsets proved insignificant in the simulations (not shown). The color bar scales of intensities for the three different $C_q$ values are not similar in order to visualize excitation of low $C_q$ values. In this figure, 2pls corresponds to the scheme presented in 6(b), 3pls corresponds to 6(d, e) and 4pls corresponds to 6(f, g).*

### 3.2.3 Magnetization Pathway Simulations

Better understanding of the magnetization pathway during the pulse schemes is obtained by calculating the value of the density matrix ρ at different times (Δt=1μs) during a particular pulse sequence. For a two-pulse excitation scheme, figure 11 shows snapshots of ρ at equilibrium, after the first pulse, after the delay, and during the second pulse-block.

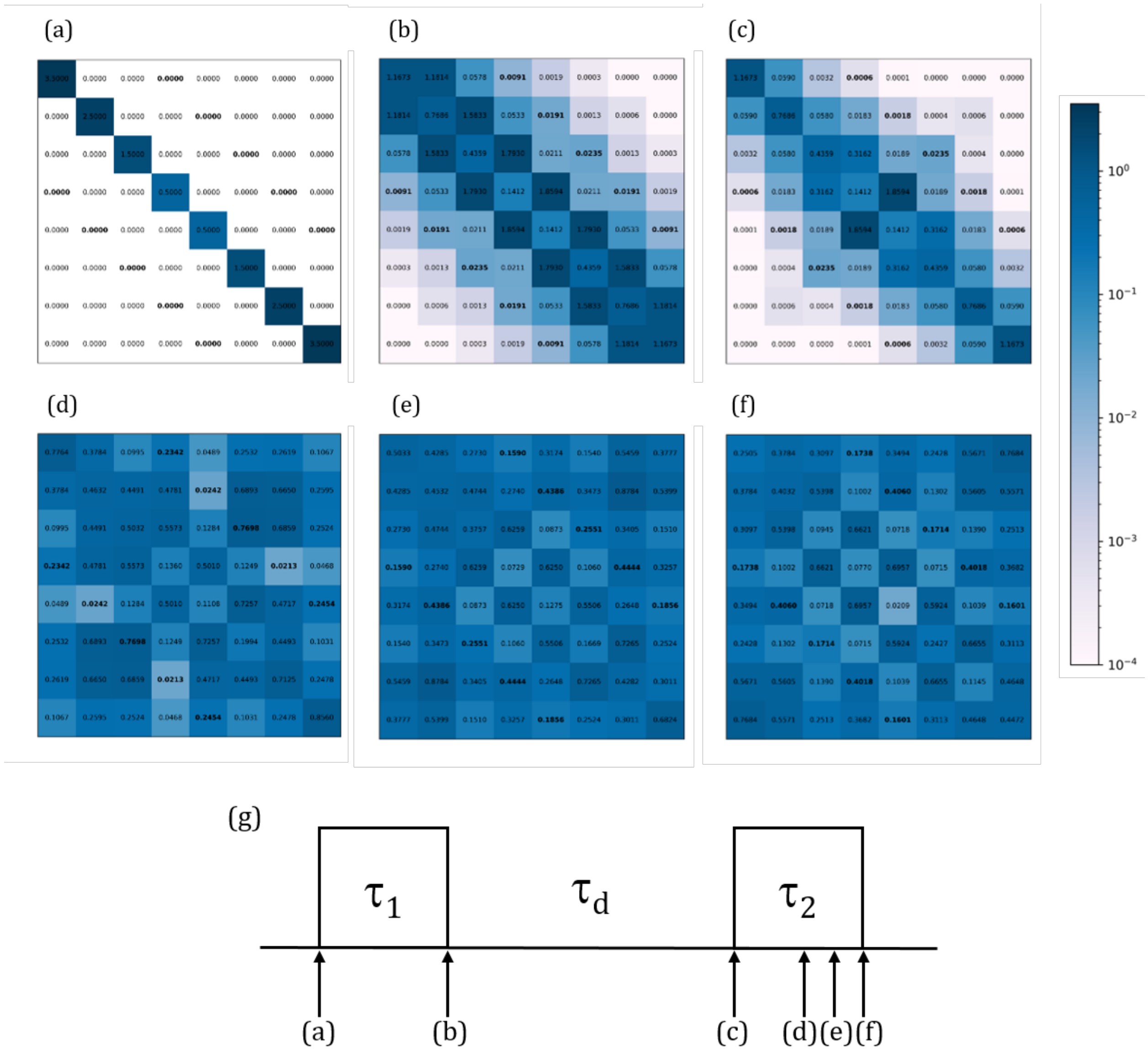

***Figure 11:*** *Absolute values of the powder-averaged density matrix during a two-pulse excitation scheme (figure 6b) of a cesium site with $C_q$=250 kHz. Here $\nu_R$=5 kHz, $\nu_1$=40 kHz. The density matrix is shown at (a) equilibrium, (b) after a 5 µs pulse, (c) after a 50 µs delay $\tau_d$ and (d), (e), (f) during the second pulse at 3, 4, and 5 µs, respectively. (g) Visualization of the sampled time points. An animation of the density matrix evolution over time is available on GitHub and in the supplementary material, video 1.1. Triple-quantum coherences are in bold.*

At equilibrium, only the zero-quantum population elements are non-zero. After the first pulse, most of the magnetization was transferred to SQC, with some very small DQC and TQC generated due to energy level mixing by the quadrupolar coupling interaction. During the delay, the different crystallites oscillate at distinctive frequencies due to the anisotropic nature of quadrupolar interaction, leading to destructive interference and therefore a reduction of the total integrated intensity of the density matrix in figure 11(b)-(f). However, the application of pulses non-synchronously recouples the quadrupolar interaction, enabling progressive excitation of higher-order coherences during the second pulse. This results in an increase of the sum of TQC elements. While the sum of TQC and 5QC is maximal in 11(e), in 11(f) the 7QC signal is maximal. It is interesting to note that during the second pulse, 5QC and 7QC are generated almost as efficiently as the TQC, although the pulse durations, phases and delays were not optimized for the excitation of these coherences. This means that the two-pulse sequence may be beneficial in excitation of higher order coherences in cesium compounds when $C_q$~250 kHz, and possibly higher. We are currently further exploring this direction.

Similar simulations were performed for a site with $C_q$=10 kHz using the two-pulse, three-pulse and the four-pulse schemes and are shown in figure 12. For this low $C_q$ value, TQC are progressively generated; however, excitation of 5QC and 7QC was negligible compared to the TQC. These simulations once again display the superiority of the four-pulse scheme over the two-pulse scheme in terms of TQC generation.

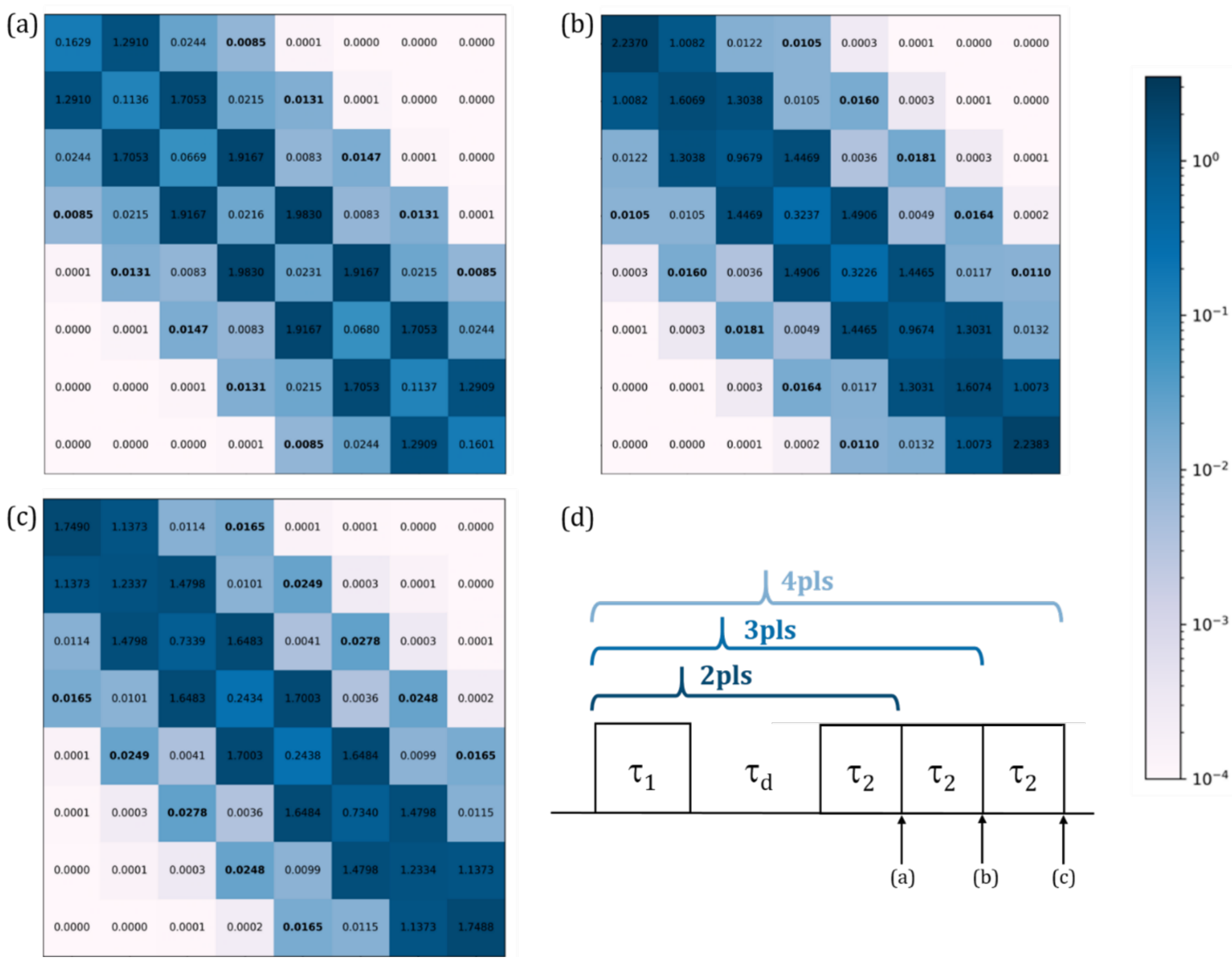

***Figure 12:*** *Absolute values of the powder-averaged density matrix during a multiple-pulse excitation scheme of a cesium site with $C_q$=10 kHz. Here $\nu_R$=5 kHz, $\nu_1$=40 kHz. The density matrix is shown in (a) after a two-pulse scheme (figure 6b), in (b) after a three-pulse scheme (figure 6d), and in (c) after a four-pulse scheme (figure 6f). An animation of the density matrix evolution over time is available on GitHub and in video 1.2 in the supplementary material. Here $\tau_1$=6 µs, $\tau_2$=3 µs, $\tau_d$=50 µs.*

## 3.3 $^{133}$Cs Triple-Quantum Spectroscopy – Experiments

### 3.3.1 triple-quantum filtered 1D $^{133}$Cs spectra

Each of the TQ filtered schemes described in figure 6 was experimentally optimized on a sample containing a mixture of Cs-loaded Zeolite X (ZX) and zeolite A (ZA). The optimal spectrum for each of the schemes, optimized separately for the ZX site (36.4 ppm) and for the ZA site (0.5 ppm), is shown in figure 13.

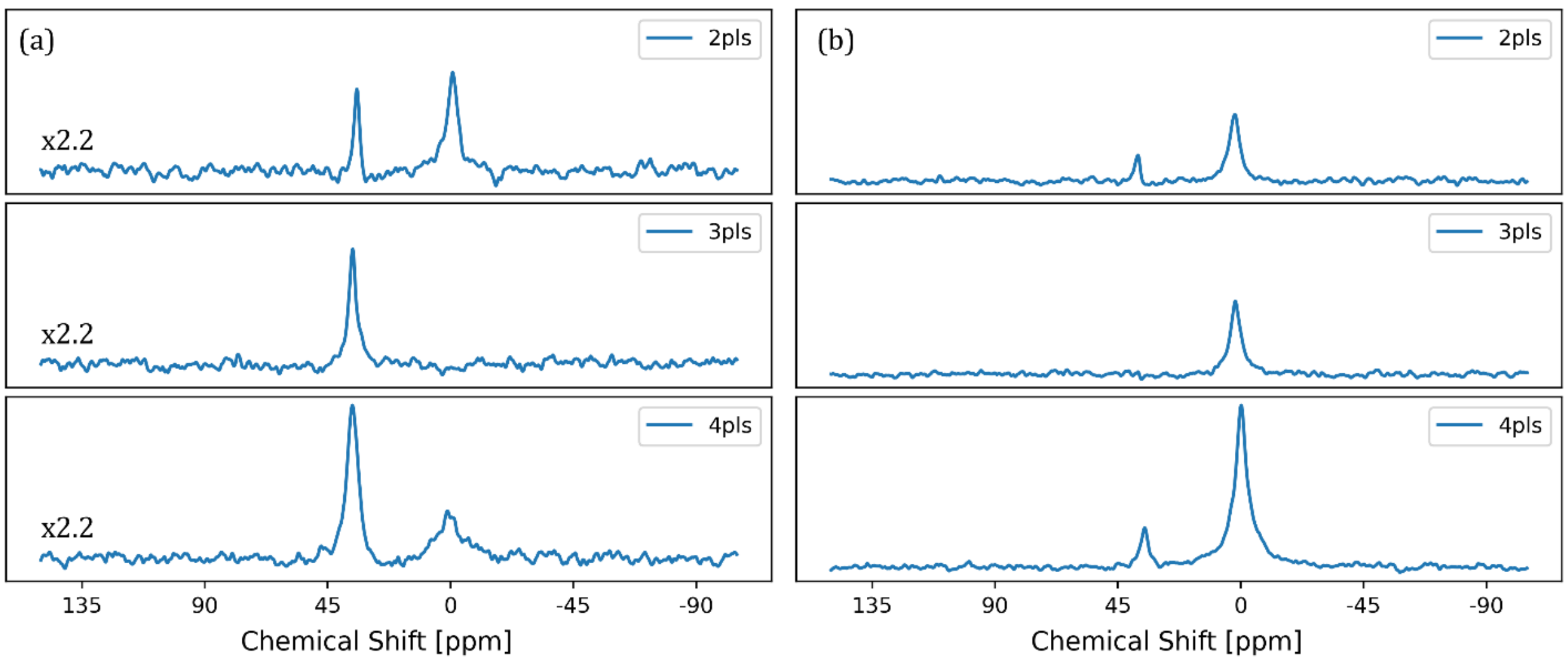

**Figure 13:** *TQ filtered signal intensity achieved by each of the schemes in figure 6. The peak at 0.5ppm belongs to ZA and the peak at 36.4ppm belongs to ZX. (a) The signal was optimized for the ZX site and normalized to the that of the four-pulse scheme, and the spectra were then magnified times 2.2 to improve visibility. (b) The signal was optimized for the ZA site and normalized to the that of the four-pulse scheme. In all experiments the number of scans was 360, $\nu_R$=5 kHz, $\nu_1$=37 kHz, and the spectra were processed with an exponential line broadening factor of 100 Hz. The pulse duration and delays are (a) two-pulse scheme: $\tau_1$=7 µs, $\tau_2$=5 µs, $\tau_d$=25 µs; three-pulse scheme: $\tau_1$=6 µs, $\tau_2$=6 µs, $\tau_d$=25 µs; four-pulse scheme: $\tau_1$=6 µs, $\tau_2$=3 µs, $\tau_d$=25 µs. (b) Two-pulse scheme: $\tau_1$=6 µs, $\tau_2$=5 µs, $\tau_d$=25 µs; three-pulse scheme: $\tau_1$=6 µs, $\tau_2$=3 µs, $\tau_d$=25 µs; four-pulse scheme: $\tau_1$=6 µs, $\tau_2$=3 µs, $\tau_d$=100 µs.*

The NMR experimental results clearly suggest that pulse-schemes that contain more pulses yield better excitation for both sites in the sample. Both the ZA and the ZX sites responded best to the four-pulse scheme showing increased peak intensities and responded most poorly to the two-pulse scheme, with 2-pulse:3-pusle:4-pulse intensity ratios of 1:1.16:1.56 for ZX and 1:1.10:2.46 for ZA. The observation that more pulses generated better efficiency in the experiment supports the results from nutation and lineshape experiments showing that the $C_q$ values of $^{133}$Cs sites in the zeolites are small, in the order of $10^1$ kHz.

Achieving efficient excitation of both sites simultaneously is shown to be challenging; both sites are highly sensitive to small changes in pulse durations or delays between the pulses (see experimental optimizations in the SI, figures S2.1, S2.2). Even a change of 1µs in the pulse length could have a dramatic impact on the relative excitation efficiency of the two sites. This is probably another manifestation of small quadrupole coupling constants causing the excitation of TQCs in spin-7/2 systems to be very challenging. Another manifestation of the effect of small changes in $C_q$ on the excitation efficiency can be found in the simulations provided in figure S3.1 where the efficiency of one particular four-pulse sequence is plotted for $C_q$ values in the range of 1-15 kHz. On the other hand, selective excitation profiles may be used for filtering sites based on the quadrupolar frequencies, which is a different means of reducing spectral complexity. If better signals and better uniformity are desired, a larger space of experimental variables must be optimized efficiently when seeking the best MQC excitation scheme. This is ongoing work in our lab.

### 3.3.2 Two-dimensional TQ/SQ $^{133}$Cs correlation spectra

Two-dimensional (2D) TQ/SQ experiments were conducted on the model mixed zeolites sample using the two- and four-pulse schemes.

The 2D spectra presented in figure 14 demonstrate that the four-pulse scheme is favorable for excitation of TQC as predicted by our simulations. Although the peak positions in the TQ dimension are dispersed by a factor of three compared to the SQ dimension, there is also a three-fold increase in line widths. Prior studies suggest that in dehydrated zeolites A and X, Cs can occupy several different sites, and that those are indistinguishable in the hydrated form due to site exchange[28–30,59–61]. The increase in linewidth observed in our TQ dimension can be attributed to several factors; inhomogeneous broadening due to the presence of water molecules, the different number of $Na^+$ ions in the vicinity of Cs, and different environments generated by the Al and Si in the zeolitic framework. Other possible sources of broadening are paramagnetic effects of oxygen molecules that scale linearly with the order of coherence[62], faster relaxation of the higher order coherences, and the chemical exchange of $Cs^+$ ions between the different cesium-binding sites[29]. It remains a challenge to determine which of these mechanisms is the dominant factor in the observed line broadening.

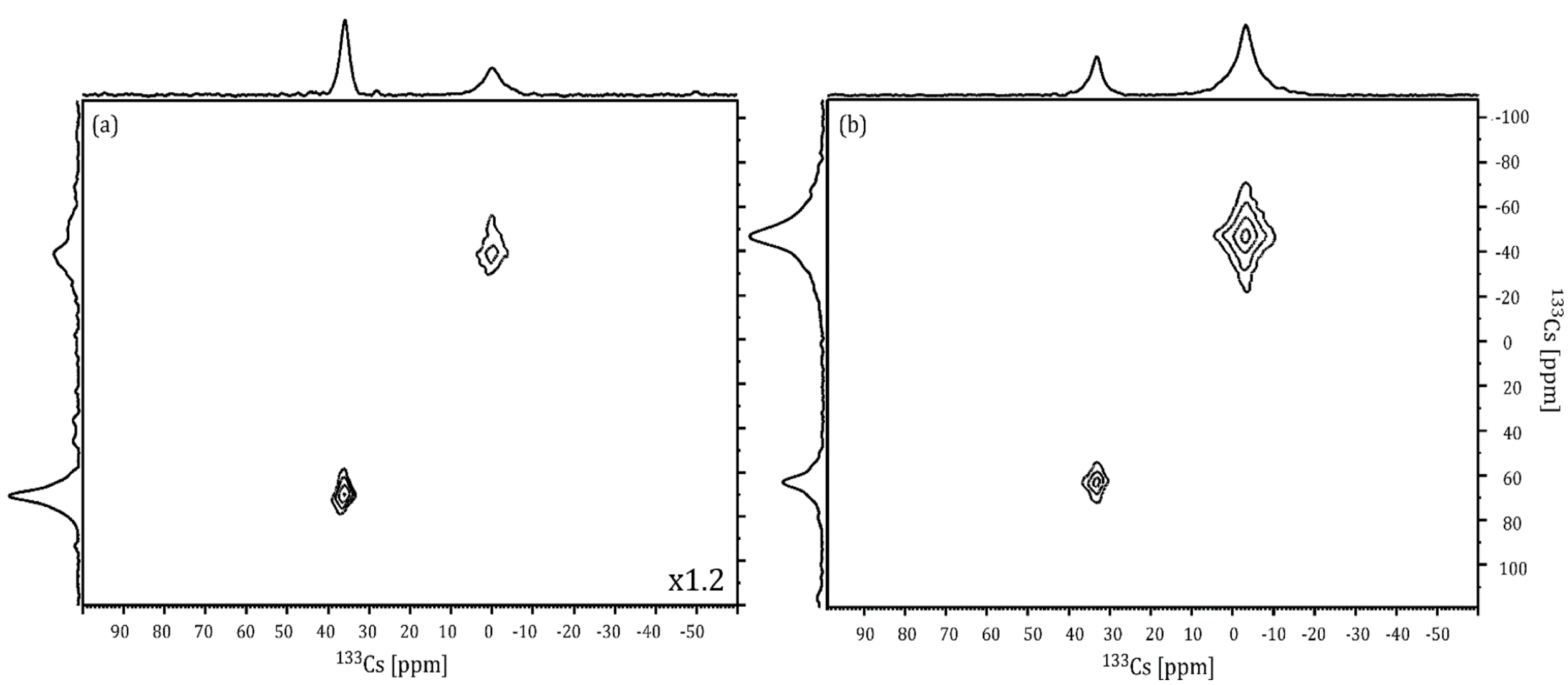

**Figure 14:** *Two-dimensional experimental* $^{133}$*Cs TQ/SQ MAS correlation spectra on the sample containing a mixture of Cs-exchanged zeolites A and X. The vertical axes represent the TQ dimension and horizontal axes represent the SQ detection dimension. The spectra were obtained using (a) a two-pulse scheme with* $\tau_1$=7 μs, $\tau_2$=5 μs, $\tau_d$=25 μs, *and (b) a four-pulse scheme with* $\tau_1$=6 μs, $\tau_2$=3 μs, $\tau_d$=100 μs. *The number of scans was 360,* $\nu_R$=5 kHz, $\nu_1$=37.5 kHz, *and the spectra were processed with an exponential line broadening factor of 100 Hz. The signal in (a) was magnified times 1.2 to visually normalize the projection peaks to that of the four-pulse scheme.*

## 3.4 Summary & Conclusions

Common pulse sequences used in multiple-quantum NMR spectroscopy utilize one-pulse, composite, or shaped-pulse techniques to excite MQCs when the quadrupolar frequency $\nu_Q$ is 0.5 MHz and above. For lower quadrupolar frequency values, two pulses separated by a delay have been shown to be a more useful way to excite MQCs. However, both methods fail to provide sufficient sensitivity for $^{133}$Cs, a spin-7/2 in the hydrated zeolite matrices studied here. This is due to the weak quadrupolar coupling constant $C_q$, below ~20 kHz, giving rise to low efficiency of multiple quantum excitation.

In order to enhance the TQC excitation efficiency, we further modified existing excitation schemes by employing three and four-pulse excitation blocks. These pulse blocks proved to increase the signal-to-noise ratio (SNR) considerably by a factor of ~2, both in simulations and, more importantly, in experiments. This enhancement in SNR, although not yet uniform and strongly dependent on $C_q$, results in efficient two-dimensional spectra that can aid in the identification of cesium-binding sites. In the case demonstrated here for hydrated Cs-exchanged zeolites A and X, the increased dispersion afforded by encoding a triple-quantum dimension did not produce actual resolution enhancement since the line widths increased three-fold, similar to the increase in peak dispersion. We conclude that although factors like paramagnetism and relaxation may contribute to this broadening, it is also reasonable to assume that Cs sites in zeolites A and X have structural heterogeneity, which may be due to several reasons; (i) the variability in the zeolitic environment containing both Si and Al, (ii) possible existence of both water and Na ions in the vicinity of neighboring Cs sites, and (iii) chemical exchange between different cation binding sites.

It is our aim to apply this improved triple-quantum excitation scheme, demonstrated here for a model of two well-defined Cs-loaded zeolites, for the characterization of Cs binding sites in more complex matrices such as zeolite-geopolymer composites, and possibly cementitious matrices. Such studies could provide better insight into cesium binding in these immobilization materials.

## 4. ACKNOWLEDGEMENTS

This research was supported by the Pazy foundation awarded to AG and MAH, grant no. (701)-2025.

# ENHANCED $^{133}$CS TRIPLE-QUANTUM EXCITATION IN SOLID-STATE NMR OF CS-BEARING ZEOLITES

N. Vaisleib[1], M. Arbel-Haddad[2], A. Goldbourt[1§]

[1]School of Chemistry, Tel Aviv University, Ramat Aviv 6997801, Tel Aviv, Israel
[2]Nuclear Research Center Negev, PO Box 9001, Beer Sheva 84901, Israel

[§]amirgo@tauex.tau.ac.il

## SUPPLEMENTARY MATERIAL

### S1: Quadrupolar Lineshapes Simulations

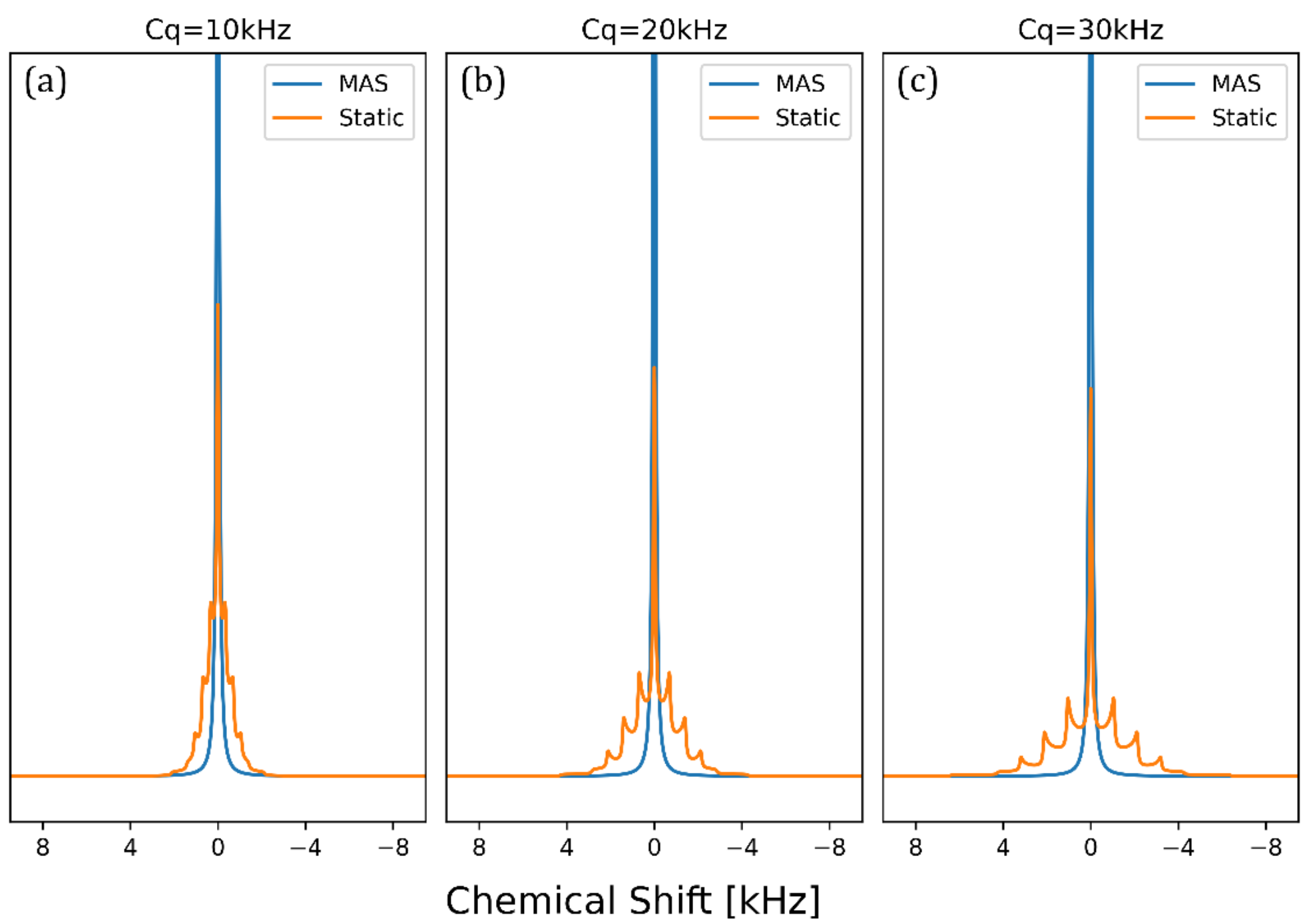

***Figure S1.1:*** *(a-c) Simulations of $^{133}$Cs MAS (blue, $\nu_R$=15 kHz) and static (orange) lineshapes for different $C_q$ values, while $\eta_q$=0. Only the first-order quadrupolar interaction was considered. In order to include ST sidebands, the detection operator was chosen to be $I^+$. Line broadening of 100 Hz was applied in all cases.*

### S2: Experimental Optimizations on $^{133}$Cs

The optimizations of $^{133}$Cs triple-quantum excitation filtered signals, shown in figure S1.1, were performed on the pulse lengths using a fixed delay of 50 μs, for pulse durations of up to 12μs. The values of $\tau_1$ and $\tau_2$ (definitions in figure 6) were incremented in steps of 1μs such that $\tau_2$ was incremented across its entire range before incrementing $\tau_1$ (nested incrementation). The delay time $\tau_d$ was fixed to 50 μs.

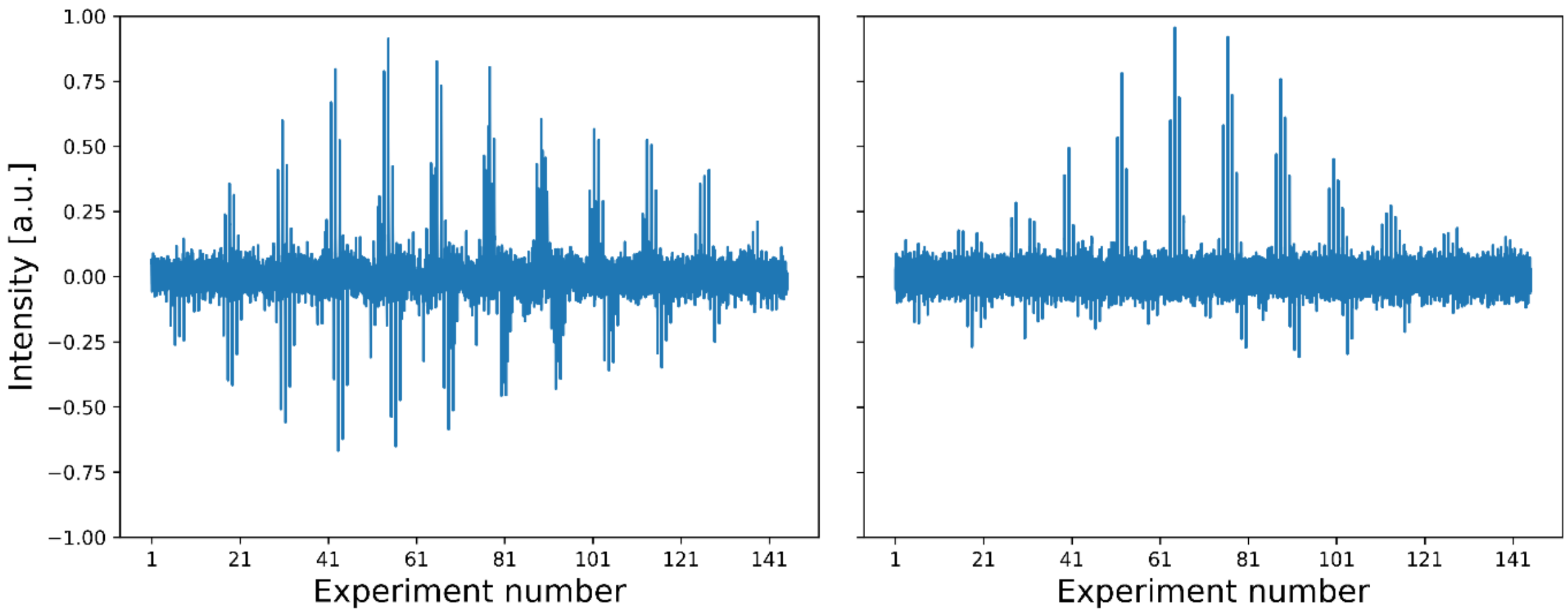

***Figure S2.1:*** *Optimizations of pulse durations for (a) two-pulse* $^{133}$*Cs triple-quantum excitation scheme and (b) three-pulse* $^{133}$*Cs triple-quantum excitation scheme (see figure 6) with a fixed delay. The signal was normalized to the maximal intensity achieved.*

The optimizations $^{133}$Cs triple-quantum excitation filtered signals shown in figure S1.2 were made for pulse durations of up to 10μs and for the delays up to 1 rotor period. The values of $\tau_1$ and $\tau_2$ (definitions in figure 6) were incremented in steps of 1μs. The delay values were incremented in steps of ¼ rotor period (25 μs). Increments in $\tau_d$, $\tau_1$ and $\tau_2$ were performed such that $\tau_d$ was incremented across its entire range before incrementing $\tau_2$, also incremented across its entire range, then $\tau_1$ (nested incrementation).

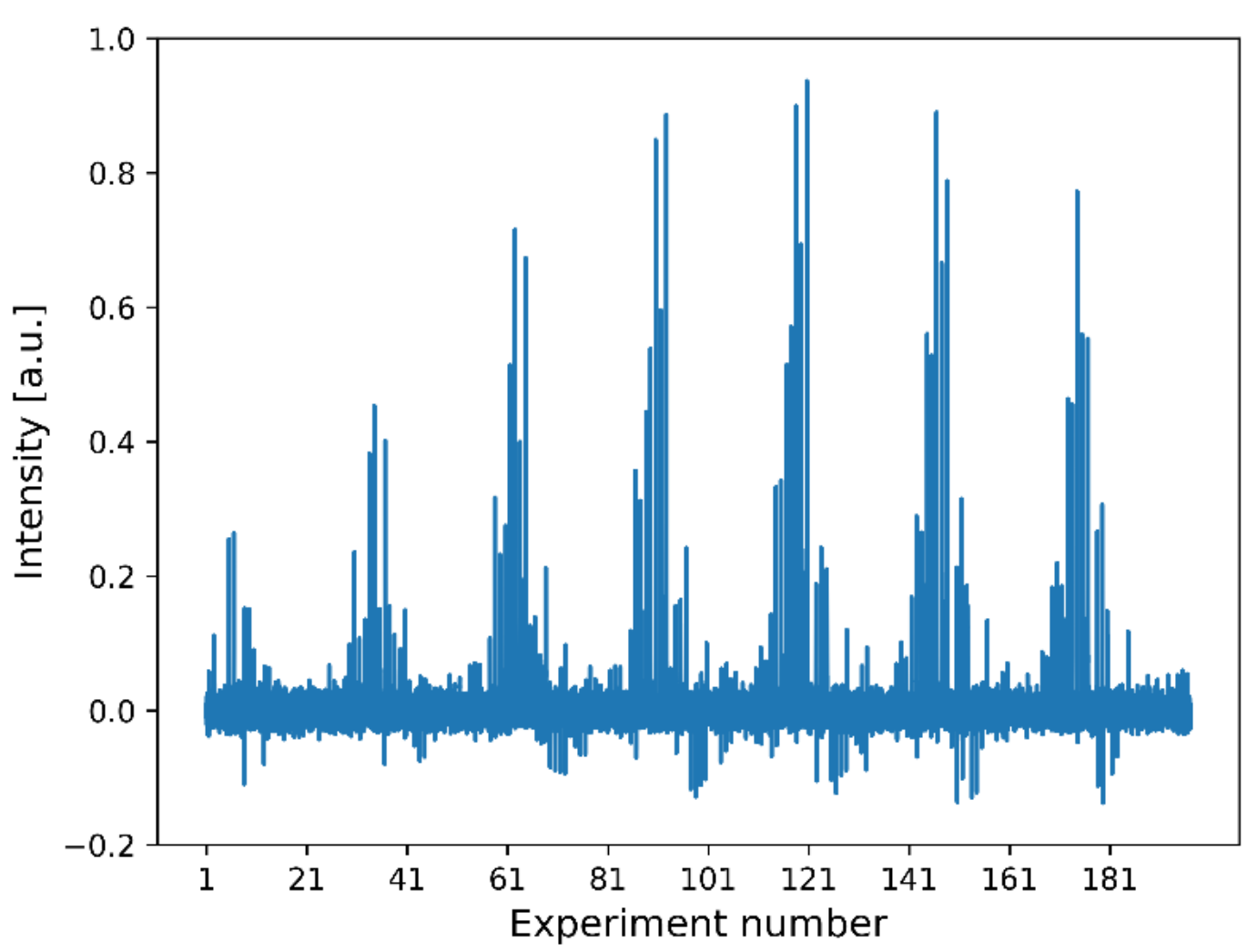

***Figure S2.2:*** *Optimizations of $^{133}$Cs triple-quantum filtered signal. Pulse durations and delays between the two pulse blocks in a four-pulse triple-quantum excitation scheme were varied as explained in the text.*

## S3: Excitation efficiency vs quadrupolar coupling constant

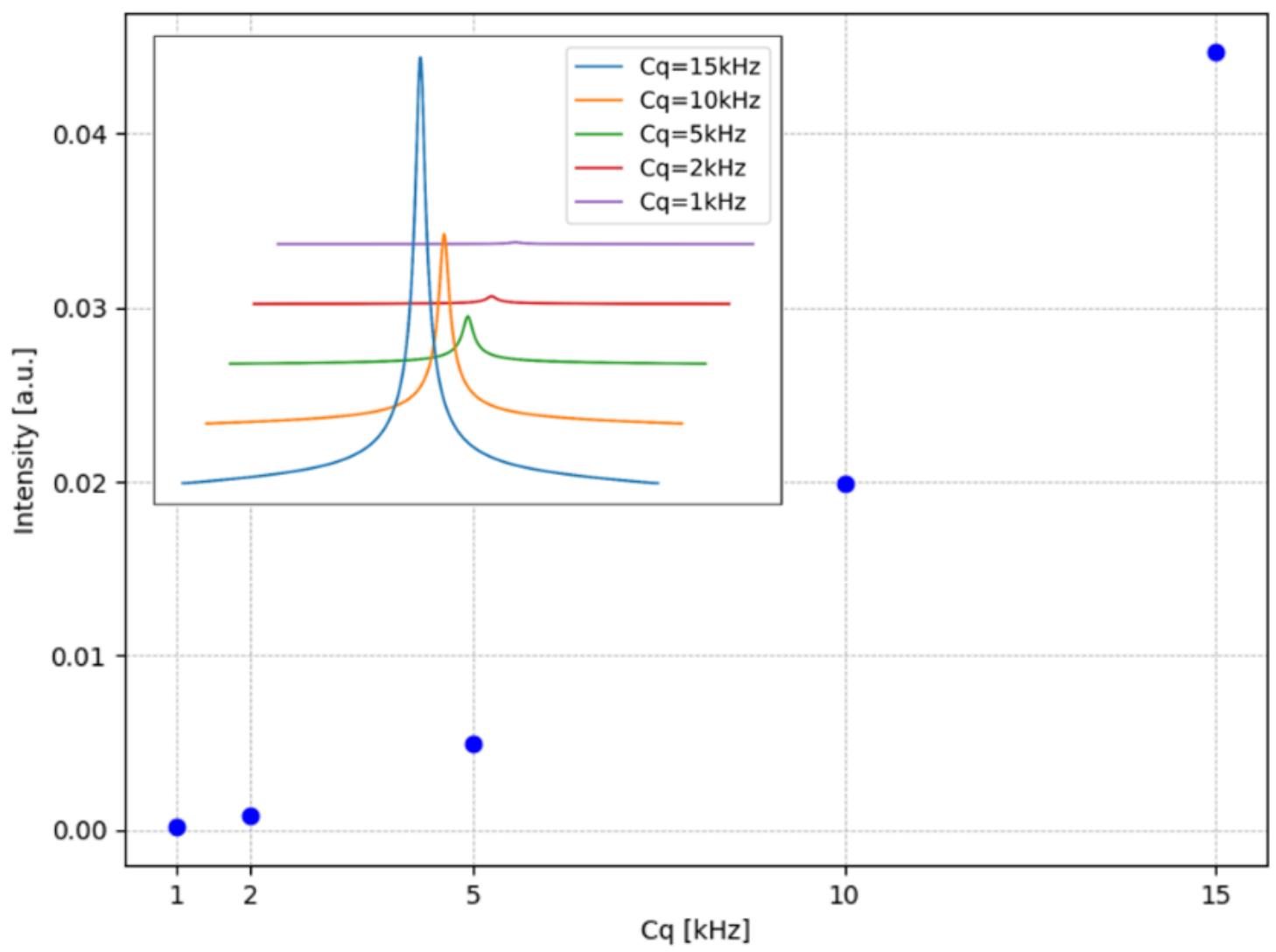

***Figure S3.1:*** *Triple-quantum excitation efficiency plotted as a function of Cq in the range of 1-15 kHz. To demonstrate the strong dependence on Cq, we used a single four-pulse block excitation scheme optimized for Cq=15 kHz with $\tau_1$=5 μs; $\tau_d$=50 μs; $\tau_2$=5 μs. Spinning speed is 10 kHz.*

## V1: Density matrix animations (also available on GitHub).

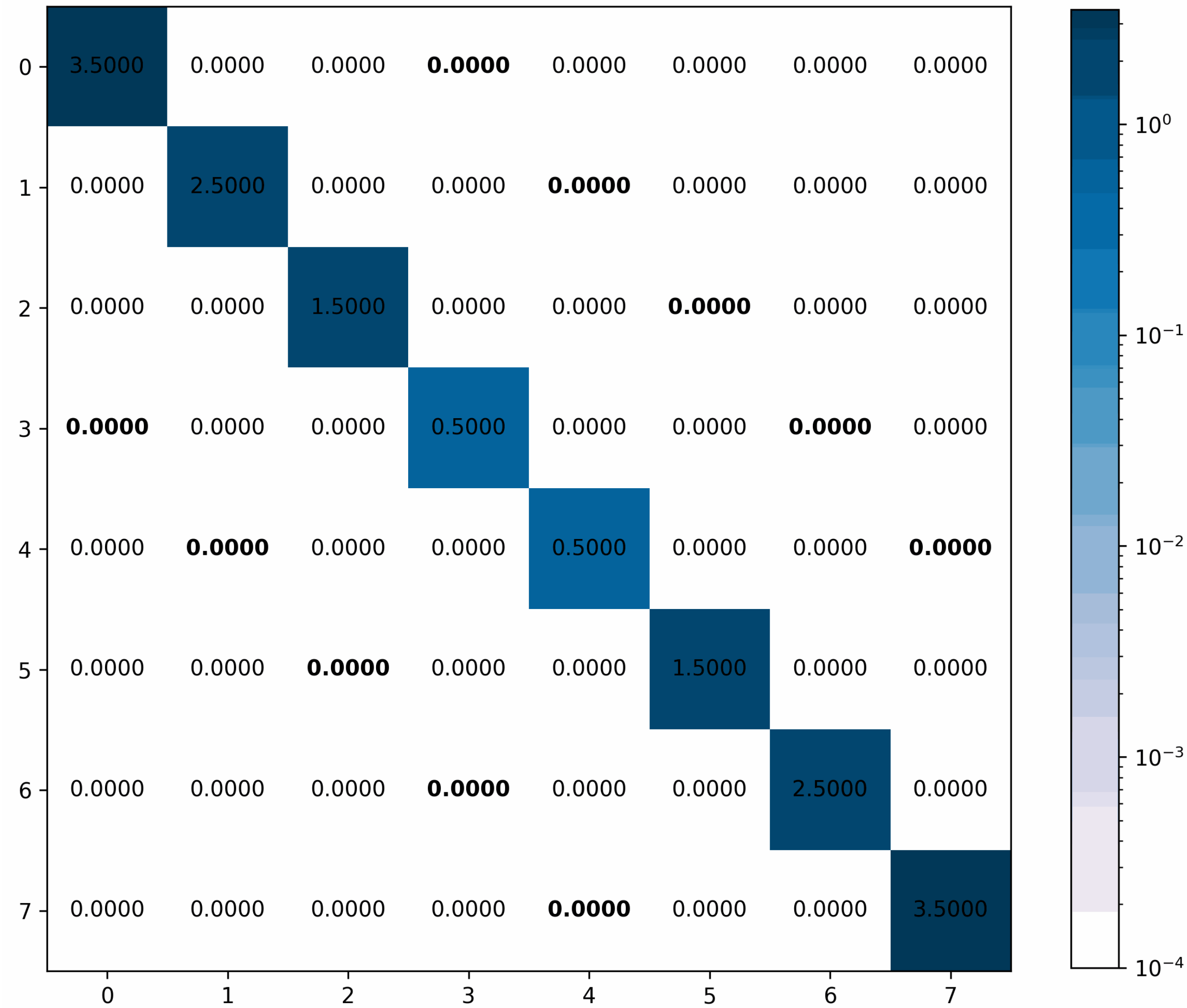

***Video 1.1:*** *The evolution of the density matrix during a two-pulse scheme. Cq=250 kHz, $\nu_R$=5 kHz, $\nu_1$=40 kHz, $\tau_1$=5 $\mu$s, $\tau_2$=5 $\mu$s, $\tau_d$=50 $\mu$s, Filename in GitHub: "250KHz-2pls.gif".*

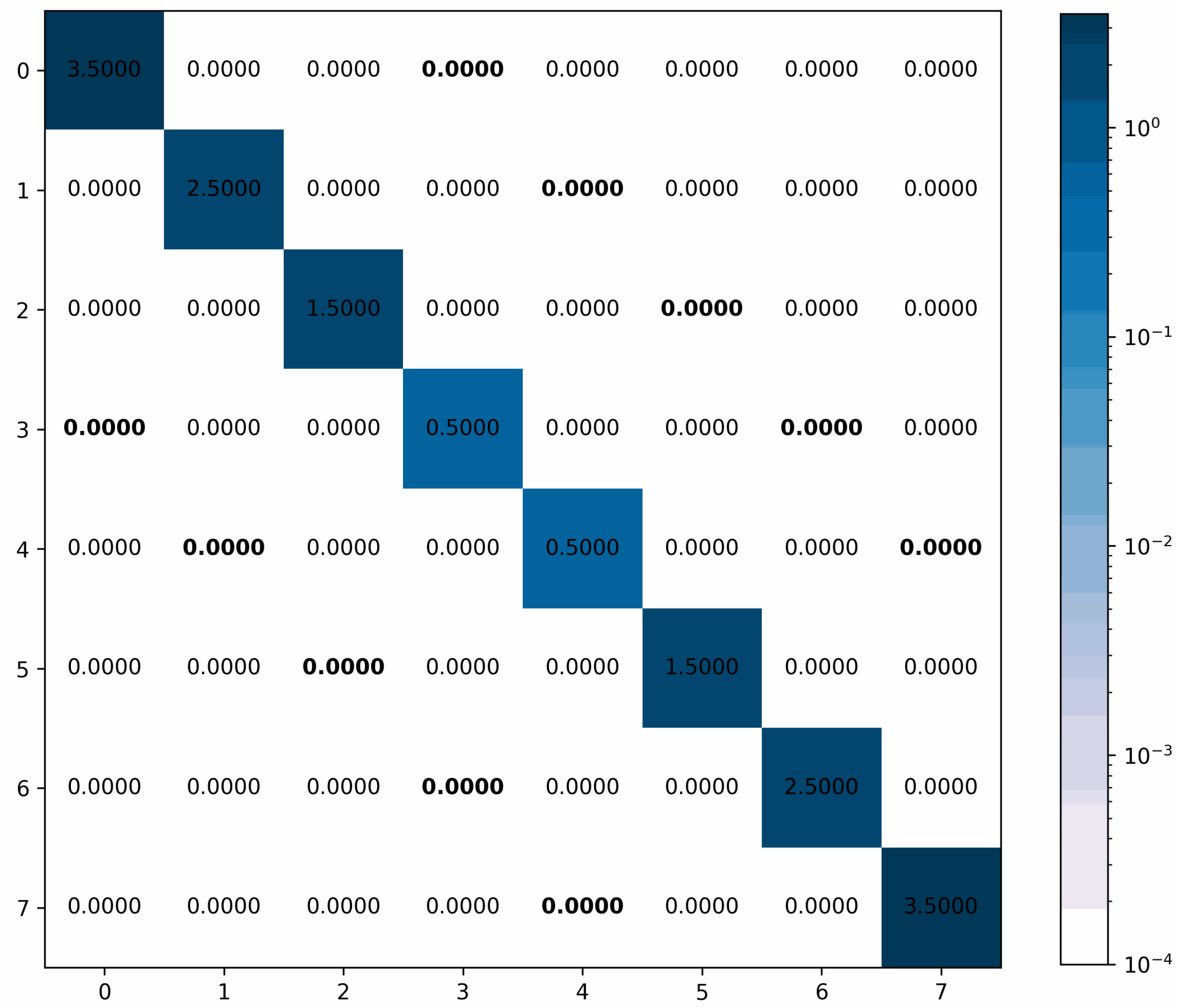

***Video 1.2:*** *The evolution of the density matrix during a four-pulse scheme. Cq=10 kHz, $\nu_R$=5 kHz, $\nu_1$=40 kHz, $\tau_1$=6 μs, $\tau_2$=3 μs, $\tau_d$=50 μs, Filename in GitHub: "10kHz-4pls.gif".*